\documentstyle[12pt,epsfig]{article}
\newfont{\Bbb}{msbm10 scaled 1200}     
\newcommand{\mathbb}[1]{\mbox{\Bbb #1}}

\def\lbldef#1#2{\expandafter\gdef\csname #1\endcsname {#2}}

\def\href#1#2{#2}

\newcommand{\beq}{\begin{equation}}
\newcommand{\eeq}{\end{equation}}
\newcommand{\ber}{\begin{eqnarray}}
\newcommand{\eer}{\end{eqnarray}}

\newcommand{\beqar}{\begin{eqnarray}}

\newcommand{\eeqar}{\end{eqnarray}}

\newcommand{\ba}{\begin{eqnarray}}
\newcommand{\ea}{\end{eqnarray}}

\newcommand{\dsl}
  {\kern.06em\hbox{\raise.15ex\hbox{$/$}\kern-.56em\hbox{$\partial$}}}

\newcommand{\eeqarr}{\end{eqnarray}}
\newcommand{\ZZ}{{\rm \kern 0.275em Z \kern -0.92em Z}\;}
\def\be{\begin{equation}}
\def\ee{\end{equation}}
\def\bea{\begin{eqnarray}}
\def\eea{\end{eqnarray}}

\begin{document}
\baselineskip=15.5pt
\pagestyle{plain}
\setcounter{page}{1}

\begin{titlepage}

\rightline{\small{\tt UB-ECM-PF-09/05 }}

\begin{center}
\vskip 1.4 cm
{\LARGE{\bf Noether symmetries, energy-momentum tensors and conformal invariance in classical field theory}}
\vskip 1.5cm

{\Large{
 Josep M. Pons}}\\
\vskip 1.2cm
\textit{ Departament ECM and  ICC, Facultat de F\'{\i}sica, Universitat de Barcelona, \\
 Diagonal 647, E-08028 Barcelona, Catalonia, Spain.}\\
\end{center}

\vspace{12pt}

\begin{center}
\textbf{Abstract}
\end{center}

\vspace{4pt}

{\noindent In the framework of classical field theory, we first review the
Noether theory of symmetries, with simple rederivations of its essential results, with
special emphasis given to the Noether identities for gauge theories. Will this baggage on board,
we next discuss in detail, for Poincar\'e invariant theories in flat spacetime, the
differences between the Belinfante energy-momentum tensor and a family of
Hilbert energy-momentum tensors. All these tensors coincide on shell
but they split their duties in the following sense: Belinfante's tensor is the one to
use in order to obtain the generators of Poincar\'e symmetries and it is a basic 
ingredient of the generators of other eventual spacetime symmetries which may happen to exist. 
Instead, Hilbert tensors are the means to test whether a theory contains
other spacetime symmetries beyond Poincar\'e. We discuss at length the case of scale and conformal symmetry, 
of which we give some examples. We show, for Poincar\'e invariant Lagrangians, that the 
realization of scale invariance selects a unique Hilbert tensor which allows for an easy 
test as to whether conformal invariance is also realized. Finally we make some basic remarks on metric 
generally covariant theories and classical field theory in a fixed curved bakground.}

\vfill \vskip 5.mm \hrule width 5.cm \vskip 2.mm {\small \noindent e-mail: pons@ecm.ub.es}

\noindent

\end{titlepage}

\newpage

\section{Introduction}
It might seem an almost impossible task to say something new as regards the theory of Noether symmetries in
classical field theory, even more so if the considerations are mostly made in flat space. Noether theory, which has 
been widely developed and employed in mathematical and theoretical physics since its dawn in 1918 \cite{noet} (for more
modern expositions of Noether theorems see \cite{Hill51}\cite{Byers98}), seems quite complete.
There are a certain number of subjects,
though, which are scarcely considered in the literature and which are nevertheless relevant in order to
extract and take advantage of all the potentialities of the Noether formulation.
We mention in this respect the detailed connection between Belinfante's \cite{Belinfante} and Hilbert's \cite{Hilbert}
approaches to the energy-momentum tensor for Poincar\'e invariant theories, first investigated by Rosenfeld \cite{Rosenfeld}, and their different role either in giving the generators of the symmetry or in implementing the conditions for the existence of a symmetry. In particular,
the role of a specific Hilbert tensor  --among a variety of Hilbert tensors-- in the implementation
of scale and conformal symmetry in flat spacetime. Relevant papers on this case of conformal symmetry are
\cite{Callan:1970ze,Polchinski:1987dy}, which are mostly devoted to its realization in
quantum field theory. As regards the classical setting,
valuable contributions to Noether theory and energy-momentum tensors include \cite{Gotay,Forger:2003ut}. Our emphasis, though, is
different, particularly regarding the different roles played by the several energy-momentum tensors and the discussion on scale and conformal invariance. We believe that our approach gives an integrated picture, always in the classical setting and in the language common to physics, which completes what it is already in the literature.

The essence of Noether theory is the connexion between continuous symmetries --of a certain type-- and conservation laws (currents and
charges) for theories whose dynamics is derived from a variational principle. Another aspect of Noether theory 
concerns the canonical formalism, where the conserved charges
associated with a Noether symmetry become, through the Poisson
bracket structure, the infinitesimal generators of the symmetry.

In this paper we will show that whereas there is a single
Belinfante energy-momentum tensor associated with a theory in flat spacetime --a theory described by a Poincar\'e
invariant Lagrangian--, which may be subject to improvements\footnote{An improvement is the
addition to the energy-momentum tensor of a functional of the fields with identically vanishing divergence.},
one does not have a unique
Hilbert tensor, but a family of them, all coinciding on shell. An additional and related remark,
already pointed out by Rosenfeld, is that Belinfante's energy-momentum tensor, which is in general only symmetric on
shell, only coincides with the Hilbert tensors on shell. These observations could be thought unremarkable
were it not for two facts. One is that it is the Belinfante tensor that contains the right information to construct
the Poincar\'e symmetry generators and, basically, other spacetime symmetries that may exist; the
other is that it is a Hilbert tensor that contains the right information to examine the eventual
implementation of these other symmetries beyond
Poincar\'e, like scale and conformal invariance. The analysis presented here relies heavily on the
Noether identities for gauge theories.

The classical realization of scale invariance takes place when the Lagrangian has no dimensional parameters. 
In such case we show how to determine a specific Hilbert tensor for the theory. The trace of this tensor is 
the divergence of a quantity that can be computed. This quantity is then used to set up a simple test as to whether 
conformal invariance can be classically realized. This test coincides with the one obtained in \cite{Callan:1970ze} 
from a different analysis, not involving the Hilbert tensor.

In the subjects we have dealt with, we have tried to be complete, at the risk of, in some sections, reobtaining well 
known results. In this case we have
tried, however, to produce new, brief, and clean presentations for old subjects, in a manner which we
think could be useful for an introduction to Noether symmetries in general and Poincar\'e symmetries in flat spacetime
in particular. Whenever we encounter spacetime transformations, either in flat spacetime or as
diffeomorphisms in generally covariant theories, we always take the active view of the transformations:
acting on the fields and leaving the coordinates unchanged. We believe that the active view is the most
efficient one, and allows the spacetime and internal symmetries to be dealt with on the same footing. In
addition, it is worth noticing that, in the canonical formalism, the action of the symmetry generators through the 
Poisson bracket corresponds to the active view.

In the present paper only the bosonic case is considered.
An extension to the spinorial case is left for future work.

The organization of the paper is as follows. In section 2 we explore the basics of continuous symmetries,
including Noether symmetries, either rigid or gauge, and conserved currents, in the Lagrangian formulation.
Some remarks are made
on the existence, in gauge theories, of first class constraints. Section 3 is devoted to theories in flat
spacetime with Poincar\'e invariance, and the relation is given between the Belinfante tensor and a family
of Hilbert tensors, and their respective roles are explained. In section 4 we discuss scale and conformal
invariance and obtain, out of the trace of a specific Hilbert tensor, an expression to check for a given scale 
invariant theory whether conformal invariance is also realized. In section 5 we briefly consider generally covariant 
theories. An appendix complements section 3.

We set the notation used in the paper for the diferent energy-momentum tensors. ${\hat T}^{\mu\nu}$ stands for the
canonical energy-momentum tensor, $T_{\!b}^{\mu\nu}$ for the Belinfante tensor, and  $T^{\mu\nu}$ for a generic
Hilbert tensor. When we need to specify a Hilbert tensor associated with some density
weights of the fields, we will write $T^{(n)\mu\nu}$.
\section{Noether symmetries}
\subsection{Some identities in the variational calculus}
\label{someid}
The variational calculus exhibits some identities which are very useful in implementing the
conditions for a continuous transformation to be a symmetry.
Let us consider a field theory, governed by an action principle
\beq{\cal S} = \int {\cal L}\,,
\label{act}
\eeq
where ${\cal L}$ is the Lagrangian density, with depends on the fields and their derivatives, in a finite number.
The equations of motion are obtained by demanding extremality under arbitrary variations of the fields,
$$\delta {\cal S} = \int \delta{\cal L} = \int [{\cal L}]_A\delta \phi^A + {\rm b.t.} = 0\,,
$$
where $\phi^A$ is a generic field or field component, $[{\cal L}]_A$ stands for the Euler-Lagrange (E-L) functional
derivative of
${\cal L}$ with respect to $\phi^A$ (we use also the notation $\frac{\delta{\cal L}}{\delta\phi^A}$), and ${\rm b.t.}$
represents boundary terms,
that is, an integration on the boundary $\partial{\cal M}$ of the manifold ${\cal M}$ where the integration in
(\ref{act}) takes place.
For arbitrary $\delta \phi^A$, except for some restrictions at the boundary that will help to make the ${\rm b.t.}$ to
vanish --and thus to make ${\cal S}$ a differentiable functional \cite{Brown:1986ed}--,
the requirement $\delta {\cal S}=0$ is equivalent to $[{\cal L}]_A=0$, which are the Euler-Lagrange equations of motion (e.o.m.).

Here we are not interested as much in the dynamics as in some properties of the variations themselves. Let us now
perform a second variation so that
$$\delta (\delta \phi^A) = \frac{\partial\delta \phi^A}{\partial \phi^B} \delta \phi^B+\frac{\partial\delta
\phi^A}{\partial\phi^B_{,\mu}}\delta \phi^B_{,\mu}+\frac{\partial\delta
\phi^A}{\partial\phi^B_{,\mu\nu}}\delta\phi^B_{,\mu\nu}+\ldots\,,$$
where $\phi^B_{,\mu}:= \partial_{\mu}\phi^B$ are derivatives with respect to the coordinates of the manifold --in a
given patch--, etc. Here and henceforth, dots as in the last equation represent obvious contributions form higher derivatives of the fields.
When acting on $\delta {\cal S}$ we may choose two ways to expand the second variation, either
\beq\delta (\delta{\cal S}) = \int \delta(\delta{\cal L}) = \int [\delta{\cal L}]_A\delta \phi^A + {\rm b.t.}\,,
\label{eq1}
\eeq
(form now on ${\rm b.t.}$ is generic for any boundary term) or
\bea
\delta (\delta{\cal S}) &=& \int\delta( [{\cal L}]_A\delta \phi^A) + {\rm b.t.}=
\int\delta( [{\cal L}]_A)\delta \phi^A + \int [{\cal L}]_A\delta(\delta \phi^A) +{\rm b.t.}
\nonumber\\ &=&
\int\delta( [{\cal L}]_A)\delta \phi^A + \int [{\cal L}]_A\Big(\frac{\partial\delta \phi^A}{\partial \phi^B}
\delta \phi^B\nonumber\\ &+&\frac{\partial\delta
\phi^A}{\partial\phi^B_{,\mu}}\delta\phi^B_{,\mu}+\frac{\partial\delta
\phi^A}{\partial\phi^B_{,\mu\nu}}\delta\phi^B_{,\mu\nu}+\ldots\Big) +{\rm b.t.}\nonumber\\ &=&
\int\delta( [{\cal L}]_A)\delta \phi^A + \int \Big([{\cal L}]_A\frac{\partial\delta \phi^A}{\partial \phi^B}
\delta \phi^B\nonumber\\ &-&\partial_{\mu}([{\cal L}]_A\frac{\partial\delta
\phi^A}{\partial\phi^B_{,\mu}})+\partial_{\mu\nu}([{\cal L}]_A\frac{\partial\delta
\phi^A}{\partial\phi^B_{,\mu\nu}})+\ldots\Big)\delta \phi^B +{\rm b.t.}\,.
\label{eq2}\eea
In both expressions (\ref{eq1}) and (\ref{eq2}) the bulk term depends on $\delta\phi$, which is an arbitrary
variation. Subtracting one from the other we get
\bea 0&=&
\int \Big(\delta [{\cal L}]_A-[\delta{\cal L}]_A +[{\cal L}]_B\frac{\partial\delta \phi^B}{\partial \phi^A}
\nonumber\\ &-& \partial_{\mu}([{\cal L}]_B\frac{\partial\delta
\phi^B}{\partial\phi^A_{,\mu}})+\partial_{\mu\nu}([{\cal L}]_B\frac{\partial\delta
\phi^B}{\partial\phi^A_{,\mu\nu}})+\ldots\Big)\delta \phi^A+{\rm b.t.}\,.
\eea
Since $\delta \phi^A$ is arbitrary, the vanishing of the bulk term implies the identities\footnote{To our
knowledge, these identities were obtained in the language of mechanics by Kiyoshi Kamimura, in the early eighties, by
direct computation, and never published. A particular case of (\ref{kiyoshi}), for variations satisfying the Noether 
condition --see below--, was written in \cite{bergmann49a}, eq (2.7). We thank D. Salisbury for pointing this out to 
us.}
\beq\fbox{$ \displaystyle
\delta [{\cal L}]_A-[\delta{\cal L}]_A +[{\cal L}]_B\frac{\partial\delta \phi^B}{\partial \phi^A}
\nonumber\\ -
\partial_{\mu}([{\cal L}]_B\frac{\partial\delta \phi^B}{\partial\phi^A_{,\mu}})+\partial_{\mu\nu}([{\cal
L}]_B\frac{\partial\delta \phi^B}{\partial\phi^A_{,\mu\nu}})+\ldots=0
\label{kiyoshi}
$}\eeq
These identities are valid for an arbitrary variation $\delta \phi$. A direct check is feasible and
straightforward, but cumbersome. What the identities do in essence is to give a computation of
the variation of the E-L derivatives, $\delta [{\cal L}]_A$, in terms of combinations --including
derivatives with respect to the coordinates of the manifold-- of the E-L derivatives themselves
plus the term $[\delta{\cal L}]_A$.
\subsection{Continuous symmetries}

An immediate application of (\ref{kiyoshi}) is to establish conditions for the existence of continuous symmetries
--which we will explore with the infinitesimal variations $\delta$. Our definition of a symmetry is simple: an
invertible map sending solutions of the e.o.m. into solutions.
So let us suppose that we have a solution $\phi_0$ (that is, $\phi^A_0, \forall  A$) of the e.o.m.,
$$[{\cal L}]{}_{\big\vert_{ \phi_0}} =0\,,
$$
and let $\phi^A_0 \to \phi^A_0+\delta\phi^A_0$
be the transformation to the new configuration\footnote{As stated in the introduction, we only consider active variations on the fields. Any
transformation of the coordinates --passive transformation-- is rewritten as an active one.}, infinitesimally close to
the old one. For it to be another solution of the e.o.m., we need
 $$[{\cal L}]_{\big\vert_{ \phi_0+\delta\phi_0}} =0\,,
$$
which can also be written, at first order in the infinitesimal parameter hidden in the variation, as
$$[{\cal L}]{}_{\big\vert_{ \phi_0}} + (\delta[{\cal L}]){}_{\big\vert_{ \phi_0}}=0\,.
$$
The first term vanishes because $\phi^A_0$ is a solution. As for the second, notice that, according to (\ref{kiyoshi}),
$$(\delta[{\cal L}]){}_{\big\vert_{ \phi_0}}= [\delta{\cal L}]{}_{\big\vert_{ \phi_0}}\,,
$$
and thus we end up with the necessary and sufficient\footnote{This sufficient condition may be further 
restricted in some cases if some boundary conditions are imposed on the acceptable solutions.} condition of
the variation $\delta\phi^A$ to define an infinitesimal symmetry:
\beq\fbox{$ \displaystyle [\delta{\cal L}]{}_{\big\vert_{ \phi_0}}
= 0 \label{gensymm} $}\eeq for any solution ${\phi_0}$.

\subsubsection{Noether Symmetries}
One strong way to ensure (\ref{gensymm}) is the adoption of the
Noether setting: require that
\beq \fbox{$ \displaystyle
\delta{\cal L} = \partial_\mu F^\mu
\label{diverg}
$}
\eeq for some
$F^\mu$, which guarantees $[\delta{\cal L}]_A=0, \forall A$,
identically. We have thus proved that $\delta{\cal L}$ being a
divergence is a sufficient condition for the variation $\delta$ to
map solutions into solutions. It is worth noticing that, unlike
the general condition of symmetry (\ref{gensymm}), the Noether
condition is a direct condition on the Lagrangian and not on the e.o.m.\,. The
Noether condition must be satisfied on and off shell.

General relativity (GR) provides with an elementary example of a symmetry which is not Noether.
The Einstein-Hilbert lagrangian ${\cal L}_{{}_{\!E\!H}}=\sqrt{|g|}R $ admits a scaling symmetry (rigid Weyl rescaling) $g_{\mu\nu} \to \lambda g_{\mu\nu}$,
under which $\sqrt{|g|}R \to \lambda^{\frac{d-2}{2}}\sqrt{|g|}R$ where $d$ is the spacetime
dimension. For $d >2$ this is not a Noether symmetry but it is still a symmetry. With $ \lambda = 1+\delta\lambda$ one
has $ \delta {\cal L}_{{}_{\!E\!H}} = \frac{d-2}{2}\, \delta\lambda\,{\cal L}_{{}_{\!E\!H}}$ and thus satisfies (\ref{gensymm}).
\vspace{4mm}

The obtention of a conserved current for a Noether symmetry is
straightforward. Condition (\ref{diverg}) can also be
written as \beq\fbox{$ \displaystyle [{\cal L}]_A\delta \phi^A +
\partial_\mu J^\mu=0 \label{noethcond} $}\eeq for some current
density $J^\mu$. This is the on shell conserved current associated
with a Noether symmetry. With some additional caveats in the case
of gauge theories\footnote{There exists a projectability issue in
going from tangent space to phase space in gauge theories, see
\cite{Pons:1999az} for the general theory and
\cite{Pons:1996pr,Pons:2003uc} for its application to generally
covariant theories.}, the spatial integration of $J^0$, when
appropriately expressed in phase space, will become the generator,
under the Poisson bracket, of the Noether symmetries.

\vspace{4mm}

As a matter of notation, whenever a Noether symmetry is realized, we will say that the theory has invariance under such symmetry. This is the meaning we attach to the concept of a Poincar\'e invariant theory, a diffeomorphism invariant theory, etc.. even though the variation of the Lagrangian under the infinitesimal Noether transformation may not vanish, being in general a divergence.

\section{Noether identities}
\label{noethidsect}
Consider a field theory described by a first order Lagrangian density
--except perhaps for a divergence term-- ${\cal L}$. We also consider the
Lagrangian having some symmetries of the type
\beq
\delta \phi^A = R^A_a \epsilon^a + R^{A\mu}_a \partial_\mu\epsilon^a\,,
\label{gaugetr}
\eeq
where $\epsilon^a$ are the infinitesimal parameters of the symmetries, with the index $a$ running over the number of
independent symmetries, and with $R^A_a,\ R^{A\mu}_a$ functions of the fields and their derivatives.
Up to now everything has been general. Now let us be specific: the case of our interest is when $\epsilon^a$ are arbitrary
functions of the coordinates; then the associated symmetries are called {\sl gauge} transformations.
We assume that these gauge transformations are of Noether type, which means that
 $\delta{\cal L}$ is a divergence.

The dependence of $\delta \phi^A$ on the arbitrary functions $\epsilon^a$ imposes that
$J^\mu$ in (\ref{noethcond}) must be of the form $J^\mu= C^\mu_a\epsilon^a$ (up to the divergence of an arbitrary
antisymmetric tensor density, as we discuss below).
Hence we find, using the arbitrariness
of the functions $\epsilon^a$, that $C^\mu_a$ can be taken so as to satisfy
\beq
C^\mu_a = - [{\cal L}]_A R^{A\mu}_a\,,\qquad \partial_\mu C^\mu_a = - [{\cal L}]_A R^A_a\,.
\label{C}
\eeq
The general solution of $J^\mu$ in (\ref{noethcond}) is $J^\mu= C^\mu_a\epsilon^a + \partial_\nu N^{\nu\mu}$, with
$N^{\nu\mu}$ an arbitrary antisymmetric tensor density (we work locally and do not consider global issues). Since the
particular solution found for $C^\mu_a$ in (\ref{C}) vanishes on shell, we infer that, up to terms vanishing on shell,
the conserved Noether currents associated with gauge symmetries are always trivial and completelly undetermined by the
formalism. In the standard spacetime setting, $\mu = (0, i)$, the values of conserved charges
$\int_{{\cal M}_t} d^{d-1}\!x J^0$ (where ${\cal M}_t$ is a spacelike slice of ${\cal M}$) will depend on $N^{i 0}$ as
a boundary integral. The  $2$-form Hodge dual to $N^{\nu\mu}$ in a standard metric theory is usually called the 
superpotential.

\vspace{4mm}

Notice that elimination of $C^\mu_a$ in (\ref{C}) yields the Noether identities\footnote{This is the contents of the 
often called Noether's second theorem.}
\beq\fbox{$ \displaystyle
[{\cal L}]_A R^A_a-\partial_\mu([{\cal L}]_A R^{A\mu}_a)=0
\label{noethid}
$}\eeq

Let us pause to reobtain these identitites in another way --not essentially different, though. Let us integrate
(\ref{noethcond}) on the manifold, and in addition consider the case when the arbitrary functions $\epsilon^a$ have
compact support. This means that all boundary terms depending on the arbitrary functions --and derivatives-- will
vanish. Then we have
\beq
0 = \int [{\cal L}]_A\delta \phi^A = \int [{\cal L}]_A(R^A_a \epsilon^a + R^{A\mu}_a \partial_\mu\epsilon^a)=
\int \Big([{\cal L}]_A R^A_a-\partial_\mu([{\cal L}]_A R^{A\mu}_a) \Big)\epsilon^a\,,
\label{noethid2}
\eeq
from which (\ref{noethid}) is obtained owing to the arbitrariness of $\epsilon^a$. Saturating (\ref{noethid}) with
$\epsilon^a$, it can be equivalently written as
\beq
[{\cal L}]_A \delta\phi^A-\partial_\mu([{\cal L}]_A R^{A\mu}_a\epsilon^a)=0\,,
\label{noethid1}
\eeq
which explicitely shows the current obtained above.

Notice that in a more general case
of a gauge symmetry, for instance
\beq
\delta \phi^A = R^A_a \epsilon^a + R^{A\mu}_a \partial_\mu\epsilon^a+ R^{A\mu\nu}_a \partial_{\mu\nu}\epsilon^a + \ldots
\label{gaugetr2der}
\eeq
(where $R^{A\mu\nu}_a$ can always be taken symmetric in the $\mu\nu$ indices) we would have obtained, for the Noether
identity,
\beq\fbox{$ \displaystyle
[{\cal L}]_A R^A_a-\partial_\mu([{\cal L}]_A R^{A\mu}_a)+\partial_{\mu\nu}([{\cal L}]_A R^{A\mu\nu}_a)+ \ldots=0
\label{noethid2der}
$}\eeq
The case with second derivatives $\partial_{\mu\nu}\epsilon^a$ takes place for instance in the Palatini formalism for GR.
\subsection{First class constraints from Noether identities}
\label{constr-noethidsect}
Observe that the quantities $C^\mu_a$ in (\ref{C}) vanish on
shell, but this is not a sufficient reason to qualify them as
constraints. Here we enter the realm of the
Rosenfeld-Dirac-Bergmann\footnote{Rosenfeld's contribution, which
has been overlooked for a long time, has recently resurfaced
thanks to the work of D. Salisbury and it is discussed in
\cite{Salisbury:2007br}.} theory of constrained systems
\cite{rosenfeld30,bergmann49a,bergbrun49,bergm3,dirac50,dirac4}.
We will not dwell in this theory\footnote{See for instance \cite{Sundermeyer:1982gv} and \cite{Henneaux:1992ig}. A brief exposition of its basics can be found in
\cite{Pons:2004pp}.} but will only borrow a few concepts from it.
To distinguish what is a constraint and what is not we must first individuate an evolution parameter
in our theory. If we are in Minkowsky spacetime or in a Lorentzian manifold, we typically
consider a coordinate $x^0=t$ (for "time") such that the equal time surfaces are spacelike. The e.o.m. are second order, and initial
conditions --positions and velocities-- are given on the initial time surface and the e.o.m.
are the differential conditions required on a solution
--which will satisfy the initial conditions by construction.
An on shell vanishing quantity will qualify as a constraint if it does not depend on the
second time derivatives. This means that such a quantity places a restriction
on the initial value problem.

Now we will prove that, assuming that the functions $R^A_a$ and $R^{A\mu}_a$ do not depend on the
second time derivatives, the quantities $C^0_a$ are actually constraints. This is the situation in general covariant, Maxwell and YM theories.

To this effect, consider the second equation in (\ref{C}) written in the form ($\mu=(0,i)$)
$$
\partial_0 C^0_a = - [{\cal L}]_A R^A_a -\partial_i C^i_a.
$$
The rhs depends at most on the second time derivatives, according the our assumptions, but
this implies, looking at the lhs, that $C^0_a$ depends at most on the first time derivatives,
which
means that it is a constraint. Thus we have proven that the special combination of the e.o.m.
given by
$$
C^0_a = - [{\cal L}]_A R^{A 0}_a
$$
is a constraint.

In addition, since these constraints participate in the $0$-component of the current $J^\mu$, they must be projectable
to phase space ($\int d^3 \!x J^0$ is the generator of the Noether symmetry, and it is always projectable) and first
class\footnote{Dirac introduced the concept of a first class function as a function whose Poisson bracket with the
constraints vanishes on the constraints' surface.} because they participate in a generator of a symmetry, which necessarily preserves the constraints.
\subsubsection{A generalization}
\label{agen}
In the more general case (\ref{gaugetr2der}) with second derivatives of $\epsilon^a$, assuming that $R^A_a$, $R^{A\mu}_a$ and $R^{A\mu\nu}_a$ depend on the
fields and their first derivatives with respect to the coordinates, we find the particular solution for
$J^\mu$ in (\ref{noethcond}) as $J^\mu = C^\mu_a\epsilon^a + \partial_\mu(C^{\mu\nu}_a\epsilon^a)$, with
\beq C^{\mu\nu}_a = - [{\cal L}]_A R^{A \mu\nu}_a,\ \
C^{\mu}_a= - [{\cal L}]_A R^{A \mu}_a + 2 \partial_\nu([{\cal L}]_A R^{A \nu\mu}_a)\,.
\label{cc}
\eeq
(Note that $C^{\mu\nu}_a$ is symmetric in ${\mu\nu}$ because $R^{A \mu\nu}_a$ is so.)
Substitution of these relations into
$[{\cal L}]_A R^{A }_a + \partial_\mu C^{\mu}_a + \partial_{\mu\nu} C^{\mu\nu}_a=0$ (which is also consequence of
(\ref{noethcond})), produces the Noether identity (\ref{noethid2der}).

Notice that (\ref{cc}) shows that $C^{\mu}_a$ and $C^{\mu\nu}_a$ vanish on shell. Let us identify, in analogy with the
previous subsection, some projectable constraints out of these objects. We continue to assume that
the E-L e.o.m. are of
second order, at least in the time derivatives. Examination of the Noether identity (\ref{noethid2der}) as regards the
presence of time derivatives shows that the combinations
$[{\cal L}]_A R^{A\, 00}_a$ and $[{\cal L}]_A R^{A\, 0}_a - 2 \partial_\mu([{\cal L}]_A R^{A\, 0\mu}_a)+
\partial_0([{\cal L}]_A R^{A\, 00}_a)$ must be constraints, that is, they can not contain second order time
derivatives, otherwise the Noether identities can not be fulfilled. In terms of the coefficients in $J^\mu$, the
constraints are $C^{00}_a$ and $C^{0}_a + \partial_0 C^{00}_a$. Let us see now the logic behind this finding. The
conserved quantity associated with the Noether transformation is
\bea
G&=& \int d^3\!x J^0 = \int d^3\!x \Big(C^{0}_a\epsilon^a + \partial_\mu(C^{\mu 0}_a\epsilon^a)\Big) =
\int d^3\!x \Big((C^{0}_a + \partial_0 C^{0 0}_a)\epsilon^a + C^{0 0}_a\dot \epsilon^a\Big)\nonumber\\
&+& {\rm \ boundary \ term}
\eea
where $\dot \epsilon^a$ stands for the time derivative of the arbitrary function $\epsilon^a$. The boundary term is
relevant --and indeed essential since the bulk term vanishes on shell-- as regards the values of conserved quantities,
but the symmetry is generated by the bulk term only. When
written in terms of canonical variables --which is always possible because $G$ is projectable to phase space--, the
generator $G$ will be expressed in terms of first class constraints and it will contain an additional piece
proportional to $\ddot \epsilon^a$ --to be able to generate the transformation (\ref{gaugetr2der}). The coefficients in
this new piece will be primary first class constraints, which are invisible in the
Lagrangian formalism because their pullback to configuration-velocity space vanishes identically. Thus we have shown
that the Lagrangian constraints $C^{0 0}_a $ are projectable from configuration-velocity space to the canonical
formalism to become secondary first class constraints, and $C^{0}_a + \partial_0 C^{0 0}_a$ to become tertiary first
class constraints. Including the primary ones, these constraints will exhaust the first class constraints of the theory, if some regularity conditions are met.
\subsubsection{An example: first class constraints for general relativity}
In GR, the infinitesimal gauge transformations are given by the Lie derivative,
$$
\delta_{\!\epsilon} g^{\mu\nu}= \epsilon^\lambda\partial_\lambda g^{\mu\nu}-
g^{\mu\lambda}\partial_\lambda\epsilon^\nu-
g^{\lambda\nu}\partial_\lambda\epsilon^\mu\,,
$$
out of which we get
$$
R^{(\mu\nu)\rho}_\sigma= -(g^{\mu\rho}\delta^\nu_\sigma+ g^{\nu\rho}\delta^\mu_\sigma)\,.
$$
With the notation ${\cal L}_{{}_{\!E\!H}}$ for the Einstein-Hilbert Lagrangian and $[{\cal L}_{{}_{\!E\!H}}]_{\mu\nu} = G_{\mu\nu}$ for the Einstein tensor density,
the constraints are
$$
C^0_\sigma = - [{\cal L}_{{}_{\!E\!H}}]_{\mu\nu} R^{(\mu\nu)0}_\sigma = 2 G_{\mu\sigma}g^{\mu 0}.
$$

Up to a factor, $g^{\mu 0}$ is the vector orthonormal to the equal time surfaces,
$$
n^\mu= - \frac{g^{\mu 0}}{\sqrt{\vert g^{00}\vert }},
$$
and therefore we have deduced that the combinations of the e.o.m.,
$$G_{\mu\sigma}n^\mu\,,
$$
are constraints, i.e., they do not depend on the second time
derivatives. These constraints are the Hamiltonian and momentum
constraints of of GR. Notice that we have obtained them form first principles, without ever
examining the e.o.m.. In the phase space picture, these constraints are secondary, the primary 
ones being the momenta conjugate to the lapse and shift variables. 

\subsection{Examples of Noether identities}
\subsubsection{The Bianchi identitites in GR}
\label{covnoeth}
Sometimes it is convenient to express (\ref{gaugetr}) with the help of the covariant derivative,
\beq
\delta_{\!\epsilon} \phi^A = U^A_\rho\epsilon^\rho + R^{A\mu}_\rho \nabla_\mu\epsilon^\rho\,,
\label{gaugetrcov}
\eeq
with $U^A_\rho=R^A_\rho -R^{A\mu}_\nu\Gamma^\nu_{\mu\rho}$, 
and the Noether identity (\ref{noethid}) becomes
\beq
[{\cal L}]_A U^A_\nu-\nabla_\mu([{\cal L}]_A R^{A\mu}_\nu)=0\,.
\label{noethidcov}
\eeq

In the case of the metric field $g^{\mu\nu}$ we have simply
$$\delta_{\!\epsilon} g^{\mu\nu} = -\nabla^\mu \epsilon^\nu -\nabla^\nu \epsilon^\mu
= -(g^{\mu\rho}\delta^\nu_\sigma + g^{\nu\rho}\delta^\mu_\sigma) \nabla_\rho \epsilon^\sigma\,,$$
which identifies $U^{\{\mu\nu\}}_\rho=0$ (now $\{\mu\nu\}$ has the role of the index $A$) and 
$$R^{\{\mu\nu\}\rho}_\sigma =- (g^{\mu\rho}\delta^\nu_\sigma + g^{\nu\rho}\delta^\mu_\sigma)\,,$$
as before. The Noether identities (\ref{noethidcov}) for GR are then
$$\nabla_\rho ([{\cal L}_{{}_{\!E\!H}}]_{\mu\nu}R^{\{\mu\nu\}\rho}_\sigma) = -2 \nabla_\rho G^\rho_{\ \sigma} =0\,,
$$ which are the doubly contracted Bianchi identitites.
\subsubsection{Noether identities for Yang-Mills coupled to gravity}
For ${\cal L}_{Y\!M}= -\frac{1}{4}\sqrt{| g |} F^a_{\mu\nu} F^{a\,\mu\nu}$ , in addition to the well known Noether 
identities associated with the internal gauge symmetry, 
$${\cal D}_{\!\mu}{\cal D}_{\!\nu}(\sqrt{|g|} F^{a\,\mu\nu})=0\,,
$$ 
we get as Noether identities for the diffeomorphism invariance,
$$
-2 {\cal D}_{\!\mu} \frac{\delta {\cal L}_{Y\!M}}{\delta g_{\mu\nu}} +  \Big({\cal D}_{\!\rho}(\sqrt{| g |}
F^{a\,\rho\mu})\Big)F^a_{\sigma\mu}g^{\sigma\nu}
=0\,,$$
where ${\cal D}_{\!\rho}$ is the total covariant derivative -including the Riemmanian connection and the $SU(N)$
connection. Actually, $ {\cal D}_{\!\mu }$ in the first term acts only as the Riemmanian covariant derivative because
there are no YM indices involved. On the contrary, in the second term, the Riemmanian connection is superfluous
because $\sqrt{| g |} F^{a\,\rho\mu}$ is an antisymmetric tensor density.

\subsubsection{Noether identities for the Palatini formalism of GR}

Also for the diffeomorphism invariance, the Palatini formalism of GR exhibits the following Noether identities
$$\frac{\delta {\cal L}}{\delta g_{\mu\nu}} \nabla_{\!\lambda} g_{\mu\nu}
- 2\nabla_{\!\mu}(\frac{\delta {\cal L}}{\delta g_{\mu\nu}}g_{\nu\lambda})-2(\nabla_{\!\nu}\frac{\delta {\cal
L}}{\delta R_{\mu\nu\rho}^{\quad\, \sigma}})R_{\mu\lambda\rho}^{\quad\, \sigma}-
2\nabla_{\!\rho}\nabla_{\!\mu}\nabla_{\!\nu}(\frac{\delta {\cal L}}{\delta R_{\mu\nu\rho}^{\quad\, \lambda}})=0\,.
$$
\section{Poincar\'e invariant theories}
Here we will consider standard flat spacetime field theories with Poincar\'e invariance
satisfying the Noether condition
(\ref{noethcond}). There is a standard route to construct the energy-momentum tensor encompassing the four currents
associated with the four translational symmetries in flat space with Cartesian coordinates. An improvement
--Belinfante-- of this tensor allows for it to provide also for an easy construction of the currents associated 
with the Lorentz symmetries.
\subsection{Belinfante's energy-momentum tensor}
\label{belinten}
Although the material is standard, here we discuss, for the sake of completeness, Belinfante's improvement, 
\cite{Belinfante}, of the canonical energy-momentum tensor for Poincar\'e invariant theories in Minkowski spacetime. 
Throughout this subsection, $\epsilon^\mu$ will stand for the infinitesimal Poincar\'e coordinate transformations. In 
Cartesian coordinates,
\beq
\epsilon^\mu = a^\mu + \omega^\mu_{\ \nu} x^\nu,\quad \omega_{\mu\nu}=-\omega_{\nu\mu} \,,
\label{epspoinc}
\eeq
with $a^\mu$ and $\omega^\mu_{\ \nu}$ infinitesimal parameters
($\omega_{\mu\nu} = \eta_{\mu\rho}\,\omega^\rho_{\ \nu}$).
In the active view of the action of Poincar\'e group, the fields transform as
\beq\delta \phi^A = \epsilon^\mu\partial_\mu \phi^A + {\cal S}^{A\mu\nu}_B \omega_{\mu\nu} \phi^B\,,
\label{deltpoinc}
\eeq
with ${\cal S}^{A \mu\nu}_B = {\cal S}^{A [\mu\nu]}_B$ being a
linear representation of the Lorentz group. Let us make contact
with the notation (\ref{gaugetr}) applied to the case of
diffeomorphism transformations and show how this object ${\cal
S}^{A \mu\nu}_B$ is recovered from it. We will do it in the case
of a vector field $A_\mu$, but the method is general for scalar, vector and tensor representations\footnote{Note however that we restrict here the covariant behaviour of the fields (scalars, vectors, tensors) for we do not allow them to be densities. For instance, our scalars here transform as $\delta \phi= \epsilon^\mu\partial_\mu \phi$ whereas a scalar density of weight $n$ transforms as $\delta \phi = \epsilon^\mu\partial_\mu \phi + n \phi \partial_\sigma\epsilon^\sigma$. This restriction will be lifted in section \ref{variety}.}. An
infinitesimal diffeomorphism acts as
 $$\delta_\epsilon A_\mu = \epsilon^\sigma\partial_\sigma A_\mu + A_\sigma \partial_\mu\epsilon^\sigma\,,
$$
thus we extract, form (\ref{gaugetr}),
$$R_{(\mu)\sigma} = \partial_\sigma A_\mu\,,\ \ R_{(\mu)\sigma}^\rho = A_\sigma\delta_\mu^\rho\,,$$
Then, for $\epsilon^\mu$ as in (\ref{epspoinc}),
$$R_{(\mu)\sigma}^\rho \partial_\rho\epsilon^\sigma =  \frac{1}{2}\Big( R_{(\mu)\sigma}^\rho\eta^{\sigma\nu}-R_{(\mu)\sigma}^\nu\eta^{\sigma\rho}
\Big)\omega_{\nu\rho} =: \frac{1}{2}
{\cal S}^{(\lambda) \nu\rho}_{(\mu)}\omega_{\nu\rho}A_\lambda\,,$$
with
\beq
{\cal S}^{(\lambda) \nu\rho}_{(\mu)} =  (\delta_\mu^\rho\eta^{\lambda\nu}-\delta_\mu^\nu\eta^{\lambda\rho})
\label{vecrep}\eeq
being the matrices of the vector representation of the Lorentz algebra. Keeping (\ref{gaugetr}) in mind, in the general case we will have
\beq
{\cal S}^{A \mu\nu}_B\phi^B
= \Big( R^{A\nu}_\sigma \eta^{\sigma\mu} - R^{A \mu }_\sigma \eta^{\sigma\nu} \Big)\,.
\label{rversuss}
\eeq

All is summarized in the decomposition into antisymmetric and symmetric components
\beq R^{A \nu }_\sigma \eta^{\sigma\mu}= \frac{1}{2}({\cal S}^{A \mu\nu}_B\phi^B  + {\cal Q}^{A \mu\nu}_B\phi^B) \,,
\label{simantisim}
\eeq
where\footnote{We use the standard notation $[\ ]$ for antisymmetry and $(\ )$ for symmetry.}
${\cal S}^{A \mu\nu}_B = {\cal S}^{A [\mu\nu]}_B$ and ${\cal Q}^{A \mu\nu}_B = {\cal Q}^{A (\mu\nu)}_B$, and the observation that only the antisymmetric term contributes to (\ref{noethcond}) when $\epsilon^\mu$ is as in
(\ref{epspoinc}). The matrices $(M^{\mu\nu})^A_B:= {\cal S}^{A \mu\nu}_B$ form a representation of the Lorentz algebra.

Taking into account that for a Poincar\'e scalar Lagrangian ${\cal L}_{\!f}$ ($f$ is for flat), $\delta  {\cal L}_{\!f} = \epsilon^\mu\partial_\mu{\cal L}_{\!f} =
\partial_\mu(\epsilon^\mu{\cal L}_{\!f})$ ($\epsilon^\mu$ is just (\ref{epspoinc})), we realize the existence of the Noether
symmetries, and (\ref{noethcond}) becomes
\beq
0 = [{\cal L}_{\!f}]_A\delta \phi^A +\partial_\mu\Big(\frac{\partial\,{\cal L}_{\!f}}{\partial\phi^A_{,\mu}} \delta \phi^A -
\delta^\mu_{\ \nu} \epsilon^\nu  {\cal L}_{\!f}     \Big) =
[{\cal L}_{\!f}]_A\delta \phi^A +\partial_\mu\Big({\hat T}^\mu_{\ \rho} \epsilon^\rho+ \frac{\partial\,{\cal
L}_f}{\partial\phi^A_{,\mu}} R^{A \nu}_\rho \partial_\nu\epsilon^\rho\Big)\,,
\label{naiveform}\eeq
where the canonical energy-momentum tensor is defined as
\beq\fbox{$ \displaystyle{\hat T}^\mu_{\ \nu}=\frac{\partial\,{\cal L}_{\!f}}{\partial\phi^A_{,\mu}}\phi^A_{,\nu}-
\delta^\mu_{\ \nu}  {\cal L}_{\!f}
\label{naive}$}
\eeq

It is possible, and highly convenient, to get rid of the first derivatives of $\epsilon^\rho$ within the current in
(\ref{naiveform}). To this end observe that, using the antisymmetry of $\omega_{\sigma\nu}$,
\bea
\frac{\partial\,{\cal L}_{\!f}}{\partial\phi^A_{,\mu}}R^{A \nu}_\rho\partial_\nu\epsilon^\rho&=&\frac{1}{2}
\frac{\partial\,{\cal L}_{\!f}}{\partial\phi^A_{,\mu}}{\cal S}^{A \sigma\nu}_B\omega_{\sigma\nu}\phi^B\nonumber\\ &=&
\frac{1}{2}\Big(\frac{\partial\,{\cal L}_{\!f}}{\partial\phi^A_{,\mu}}{\cal S}^{A \sigma\nu}_B
+\frac{\partial\,{\cal L}_{\!f}}{\partial\phi^A_{,\sigma}}{\cal S}^{A \nu\mu}_B
+\frac{\partial\,{\cal L}_{\!f}}{\partial\phi^A_{,\nu}}{\cal S}^{A \sigma\mu}_B
 \Big)\omega_{\sigma\nu}\phi^B\nonumber\\&=:& F^{\nu\mu\sigma}\omega_{\sigma\nu}
\label{deff}
\eea
with $F^{\nu\mu\sigma}=F^{\nu[\mu\sigma]}$. Thus
we have, for (\ref{naiveform})
\bea 0 &=&[{\cal L}_{\!f}]_A\delta \phi^A
+\partial_\mu\Big({\hat T}^\mu_{\ \rho} \epsilon^\rho +
F^{\nu\mu\sigma}\omega_{\sigma\nu} \Big)=[{\cal L}_{\!f}]_A\delta \phi^A
+\partial_\mu\Big({\hat T}^\mu_{\ \rho} \epsilon^\rho -
F^{\nu\mu\sigma}\eta_{\nu\rho}\partial_\sigma\epsilon^\rho\Big)\nonumber\\
&=& [{\cal L}_{\!f}]_A\delta \phi^A +\partial_\mu\Big(({\hat T}^\mu_{\
\rho} + \partial_\sigma
F^{\nu\mu\sigma}\eta_{\nu\rho})\epsilon^\rho\Big)\,,
\label{belinform}
\eea
where in the last equality we have used that
$F^{\nu\mu\sigma}$ is antisymmetric in its last two indices. In
the last line of (\ref{belinform}) we identify Belinfante's
improvement of the canonical energy-momentum tensor,
\beq\fbox{$
\displaystyle
T_{\!b\ \rho} ^\mu := {\hat T}^\mu_{\ \rho} +
\partial_\sigma F_A^{\nu\mu\sigma}\eta_{\nu\rho}
\label{belinfdef}
$}
\eeq

Note that (\ref{belinform}) can be
written as
\beq
0 =[{\cal L}_{\!f}]_A\delta \phi^A
+\partial_\mu\Big(T_{\!b\ \rho}^\mu a^\rho +
\frac{1}{2}(T_{\!b}^{\mu\rho} x^\sigma-T_{\!b}^{\mu\sigma}
x^\rho)\omega_{\rho\sigma} \Big)\,,
\label{belinform2}
\eeq
which
allows for a neat identification of the currents associated with
translations and Lorentz transformations. The on shell
conservation of these currents guarantees that Belinfante's
energy-momentum tensor is symmetric on shell. In fact, using $\delta \phi^A$ from (\ref{deltpoinc}), and the
determination, from (\ref{belinform2}), of $\partial_\mu
T_{\!b}^{\mu\nu}= - [{\cal L}_{\!f}]_A\phi^A_{,\nu}$, one can compute,
also from (\ref{belinform2}), the antisymmetric component in
$T_{\!b}^{\mu\nu}$. One finds
$$T_{\!b}^{\mu\nu}-T_{\!b}^{\nu\mu} =  [{\cal L}_{\!f}]_A {\cal S}^{A \mu\nu}_B \phi^B\,.
$$
In particular we note that
$T_{\!b}^{\mu\nu} + \frac{1}{2}[{\cal L}]_A {\cal S}^{A \nu\mu}_B \phi^B
$
is symmetric on and off shell. It is also symmetric, noticing (\ref{simantisim}), the combination
$$T_{\!b}^{\mu\nu} + [{\cal L}_{\!f}]_AR^{A\ \mu }_\sigma \eta^{\sigma\nu}\,.
$$
We anticipate that this is indeed a Hilbert energy-momentum tensor, the objects to which we devote the next section.
\subsection{Hilbert's energy-momentum tensor(s)}
\label{rosenfeld}
Here we show a way to bypass the constructions made above by using Hilbert's prescription of substituting the
Minkowski metric $\eta$ by a general metric and extracting the energy-momentum tensor as the functional derivative of
the Lagrangian with respect to the new metric, which, after derivation, is set again to be Minkowski.
So consider a Poincar\'e invariant Lagrangian ${\cal L}_{\!f}(\phi,\partial\phi)$ in Minkowski spacetime.
We will assume that it is possible to "covariantize" it, that is, we assume that
there exists a scalar density Lagrangian ${\cal L}_{\!g}$, with the metric $g$ as a new field such that it becomes the
original Lagrangian ${\cal L}_{\!f}$ when $g\to\eta$. This is easy to do, by just promoting Poincar\'e scalars, vectors, tensors to scalars, vectors, tensors under diffeomorphsms and to place the usual square root of the determinant of the metric as a factor in the Lagrangian in order to make it the desired scalar density -notice that fermions are excluded as of now-. Thus we assume that, under variations given by the Lie derivative (that is, infinitesimal diffeomorphisms),
$$\delta_{\!\epsilon} {\cal L}_{\!g} = \partial_\mu(\epsilon^\mu{\cal L}_{\!g})\,.
$$

Let us define the tensor density
\beq
T_{\!g}^{\mu\nu} := -2 \frac{\delta {\cal L}_{\!g}}{\delta g_{\mu\nu}}\,.
\label{deftr}
\eeq
The Noether identity (\ref{noethid1}) associated with diffeomorphisms, now written for fields $\phi^A,\ g_{\mu\nu}$, becomes
\beq
[{\cal L}_{\!g}]_A \delta_{\!\epsilon}\phi^A-\partial_\mu\Big([{\cal L}_{\!g}]_A R^{A\mu}_\rho\epsilon^\rho - T_{\!g}^{\mu\sigma} g_{\sigma\rho}\epsilon^\rho\Big) -\frac{1}{2} T_{\!g}^{\mu\nu} \delta_{\!\epsilon}g_{\mu\nu} =0\,.
\label{noethidRos}
\eeq
Now take the limit $g\to \eta$, which implies $[{\cal L}_{\!g}]_A\to [{\cal L}_{\!f}]_A$, and define the Hilbert tensor $T^{\mu\nu}$ as the limit of $T_{\!g}^{\mu\nu}$ for $g\to \eta$. In this limit, $\delta_{\!\epsilon}g_{\mu\nu}= \nabla_\mu \epsilon_\nu+\nabla_\nu \epsilon_\mu \to \partial_\mu \epsilon_\nu+\partial_\nu \epsilon_\mu$, with $\epsilon_\mu$ now simply $\epsilon_\mu=\eta_{\mu\nu}\epsilon^\nu$, and these $\epsilon^\nu$ still being the components of an arbitrary vector field. Thus we get, in flat space, the identity
\beq\fbox{$ \displaystyle
[{\cal L}_{\!f}]_A \delta_{\!\epsilon}\phi^A-\partial_\mu\Big([{\cal L}_{\!f}]_A R^{A\mu}_\rho\epsilon^\rho - T^{\mu\sigma} \eta_{\sigma\rho}\epsilon^\rho\Big) -\frac{1}{2} T^{\mu\nu} (\partial_\mu \epsilon_\nu+\partial_\nu \epsilon_\mu) =0
\label{noethidRos2}
$}\eeq
for arbitrary $\epsilon^\nu$.
Now we can consider several cases. Let us start with the
Poincar\'e symmetry, sitting with Cartesian coordinates. Let $\epsilon^\mu$ be
as in (\ref{epspoinc}), that is, a Poincar\'e infinitesimal coordinate
transformation. The important fact is that with this
$\epsilon^\mu$, $\delta \eta_{\mu\nu} =\partial_\mu \epsilon_\nu
+\partial_\nu \epsilon_\mu=0$ (our choice for $\epsilon^\mu$
is the general solution of the equation for Killing vectors in flat space,
$\partial_\mu \epsilon_\nu+\partial_\nu \epsilon_\mu=0$). The last term in the lhs of (\ref{noethidRos2}) disappears and we end up with
\beq
[{\cal L}_{\!f}]_A \delta_{\!\epsilon}\phi^A+\partial_\mu\Big(  (T^{\mu}_{\ \rho}-[{\cal L}_{\!f}]_A R^{A\mu}_\rho) (a^\rho + \omega^\rho_{\ \nu} x^\nu)\Big) =0\,,
\label{noethidRos3}
\eeq
where now $\delta_{\!\epsilon}\phi^A$ are the infinitesimal Poincar\'e transformations of the fields or field compoments
$\phi^A$. Having (\ref{noethcond}) in mind, (\ref{noethidRos3}) identifies the Noether conserved current associated with Poincar\'e invariance as
\beq
J^\mu = (T^{\mu}_{\ \rho}-[{\cal L}_{\!f}]_A R^{A\mu}_\rho) (a^\rho + \omega^\rho_{\ \nu} x^\nu)\,.
\label{jfromt}
\eeq
As a matter of fact, $T^{\mu}_{\ \rho}-[{\cal L}_{\!f}]_A R^{A\mu}_\rho$ is exactly the Belinfante improved energy-momentum tensor obtained in section \ref{belinten}. This result, first obtained in \cite{Rosenfeld}\footnote{Eqivalent results are obtained in different contexts in many places, like in  \cite{Gotay,Forger:2003ut,Hehl:1994ue,Julia:1998ys,Borokhov:2002mk}, mostly to emphasize that both tensors coincide on shell.}, is proven in detail in the appendix. A possible ambiguity of the covariantization procedure, discussed in the next subsection, will not make any diference on this result.

\vspace{4mm}

Before moving on, let us note that it is from the Belinfante energy-momentum
tensor in (\ref{jfromt}) that one gets the generator of the Poincar\'e symmetries, as the spatial integration of the appropriate expression in phase space of the time component $J^0$ of the current, see also (\ref{belinform2}). The term $[{\cal L}_{\!f}]_A R^{A\mu}$, though vanishing on shell, is crucial to produce the right transformation\footnote{The relevance of terms vanishing on shell should be obvious after observing, form the discussion in section \ref{noethidsect}, that the generators of the gauge transformations in gauge theories are first class constraints --which vanish on shell.}.

\subsection{Covariantization procedures. Variety of Hilbert tensors}
\label{variety}
There is in principle total freedom in the choice of the density behaviour of the fields. The minimal procedure of covariantization is to promote all the fields to geometric entities without implementing any density behaviour for them. This is what we have done above. But one could also have promoted the Poincar\'e fields to field densities of arbitrary weights. The only consistency needed is that ${\cal L}_{\!g}$ be a scalar density of weight $1$, which is easily met. To every different covariantization there will correspond a different Hilbert tensor. 

Let us write $ {\cal L}_{\!g}^{(0)}$ for the minimal covariantization of $ {\cal L}_{\!f}$, that is, with all fields without density behaviour. There is a specific role of $t:=\sqrt{| g |}$ to make $ {\cal L}_{\!g}^{(0)}$ a density. Now let us introduce densitizations. Let $n_{{}_{\!A}}$ be the weight associated with $\phi^A$, so that $\tilde \phi^A:=t^{-n_{{}_{\!A}}} \phi^A$ (no sum for $A$) is not a density scalar, density tensor... any more, but just a scalar, tensor... . Let $n$ symbolize the full set of densitites $n_{{}_{\!A}}$. One can define a new scalar density Lagrangian
$$ {\cal L}_{\!g}^{(n)} = {\cal L}_{\!g}^{(0)}{}_{\Big|_{\phi^A\to\tilde \phi^A}}\,.
$$

We do not claim that this procedure will exhaust the ways of obtaining different covariantized Lagrangians but, looking at the result, it is conceivable that the Hilbert tensors thus obtained are the most general ones, except for improvements\footnote{Improvements that can also be introduced, see subsection \ref{impr}, by adding to the covariantized Lagrangian terms that vanish in flat space. An example of this procedure is given in \cite{Deser:1970hs}.}.
Noticing that $\frac{\partial\,  t}{\partial\, g^{\mu\nu}} = - \frac{t}{2}g_{\mu\nu} $ and $\frac{\partial \, \partial_\rho t}{\partial \partial_\sigma g^{\mu\nu}} = - \frac{t}{2}\delta^\sigma_\rho\, g_{\mu\nu} $, we can compute
\bea
\frac{\delta {\cal L}_{\!g}^{(n)}}{\delta g^{\mu\nu}}{\Big|_{g\to\eta}} &=& \frac{\delta {\cal L}_{\!g}^{(0)}}{\delta g^{\mu\nu}}{\Big|_{g\to\eta}} +  \Big( \frac{\partial {\cal L}_{\!g}^{(0)}}{\partial \phi^A} \frac{\partial(t^{-n_{{}_{\!A}}} \phi^A)}{\delta t} \frac{\partial t}{ \delta g^{\mu\nu}}\Big){\Big|_{g\to\eta}}\nonumber\\ &+&
\Big( \frac{\partial {\cal L}_{\!g}^{(0)}}{\partial \phi^A_{,\rho}} \frac{\partial\, \partial_\rho(t^{-n_{{}_{\!A}}} \phi^A)}{\partial t} \frac{\partial t}{ \partial g^{\mu\nu}}\Big){\Big|_{g\to\eta}}+\Big( \frac{\partial {\cal L}_{\!g}^{(0)}}{\partial \phi^A_{,\rho}} \frac{\partial\, \partial_\rho(t^{-n_{{}_{\!A}}} \phi^A)}{\partial\,\partial_\sigma t} \frac{\partial\,\partial_\sigma  t}{ \partial g^{\mu\nu}}\Big){\Big|_{g\to\eta}}\nonumber\\ &-&
\partial_\rho \Big(   \frac{\partial {\cal L}_{\!g}^{(0)}}{\partial\,\partial_\sigma \phi^A} \frac{\partial\, \partial_\sigma(t^{-n_{{}_{\!A}}} \phi^A)}{\partial\, \partial_\lambda t} \frac{\partial\,\partial_\lambda t}{ \partial\, \partial_\rho g^{\mu\nu}}\Big){\Big|_{g\to\eta}}\nonumber\\&=&
\frac{1}{2} T_{\mu\nu}^{(0)} + \frac{1}{2} n_{{}_{\!A}} \eta_{\mu\nu} [{\cal L}_{\!f}]_A \phi^A\,,
\eea
(we have used $T_{\mu\nu} = 2 \frac{\delta {\cal L}_{\!g}}{\delta g^{\mu\nu}}{\Big|_{g\to\eta}}\,,
$ and put a superscript ${}^{(0)}$ as a reminder) and so the new Rosenfend tensor $T_{\mu\nu}^{(n)}$ is
\beq\fbox{$ \displaystyle
T_{\mu\nu}^{(n)}= T_{\mu\nu}^{(0)} +n_{{}_{\!A}}  [{\cal L}_{\!f}]_A \phi^A\,\eta_{\mu\nu}
\label{newros}
$}\eeq

This result is fully consistent with what should have been expected on the grounds that if $\delta_\epsilon^{(0)}
\phi^A$ is the Lie derivative of $\phi^A$ considered without density behaviour, then
\beq
\delta_\epsilon^{(n)} \phi^A = \delta_\epsilon^{(0)} \phi^A + n_{{}_{\!A}}\phi^A\partial_\rho\epsilon^\rho\,,
\label{newtr}
\eeq
is the Lie derivative for an object of density weight $n_{{}_{\!A}}$. Notice that
$\delta_\epsilon^{(0)} \phi^A\to \delta_\epsilon^{(n)} \phi^A$ implies a change
$R^{(0)A\mu}_\rho\to R^{(n)A\mu}_\rho=R^{(0)A\mu}_\rho + n_{{}_{\!A}}\phi^A\delta_\rho^\mu$.

Using (\ref{newros}) and (\ref{newtr}) and taking into account (\ref {gaugetr}), one can easily pass form
(\ref{noethidRos2}) with superscripts ${}^{(0)}$ to the same relation with superscripts ${}^{(n)}$. Note in particular
that the Belinfante tensor
\beq\fbox{$ \displaystyle
T^{\mu}_{\!b\, \rho}=T^{(n)\mu}_{\ \rho}-[{\cal L}_{\!f}]_A R^{(n)A\mu}_\rho=T^{(0)\mu}_{\ \rho}-[{\cal L}_{\!f}]_A 
R^{(0)A\mu}_\rho
\label{tbr}
$}\eeq
still remains the same no matter which superscripts are used. The final form of (\ref{noethidRos2}) is then,
\beq\fbox{$ \displaystyle
[{\cal L}_{\!f}]_A \delta_{\!\epsilon}\phi^A+\partial_\mu ( T^{\mu}_{\!b\, \nu}\,\epsilon^\nu) - T^{\mu}_{\ \nu}\, 
\partial_\mu \epsilon^\nu =0\,
\label{noethidRos4}
$}\eeq
which is an identity in flat spacetime but for arbitrary $\epsilon^\nu$. 
Note that both the Belinfante tensor and the Hilbert tensor
have a role in this expression (\ref{noethidRos4}). A change
$T^{(0)\mu}_{\ \nu}\to T^{(n)\mu}_{\ \nu}$ in (\ref{noethidRos4}) is compensated by a change in the transformation 
$\delta_\epsilon^{(0)} \phi^A\to \delta_\epsilon^{(n)} \phi^A$, (\ref{newtr}).

\vspace{4mm}

As regards Poncar\'e symmetries, (\ref{epspoinc}), the term $\partial_\rho\epsilon^\rho$ in (\ref{newtr}) vanishes
because of the antisymmetry of $\omega_{\mu\nu}$, and thus $\delta_\epsilon^{(n)} \phi^A$ coincides with
$\delta_\epsilon^{(0)} \phi^A$, as Poincar\' e symmetries are concerned. But the tensors
$T_{\mu\nu}^{(n)}$ are different if the weights are
different. In particular, for a specific choice ${}^{(n)}$ of density weights --which means a choice of the
trasformations (\ref{newtr}) as well-- one may have other symmetries, for instance conformal invariance, which
will only hold for this particular choice. Although their difference vanishes on shell, $T_{\mu\nu}^{(n)}$ is not an
improvement of $T_{\mu\nu}^{(0)}$, but a basically different Hilbert tensor, which can be used to check 
whether a new symmetry is realized.

\vspace{4mm}

We conclude that whereas there is a unique --up to improvements-- Belinfante energy-momen\-tum tensor associated 
with a Poincar\'e invariant theory, one does not have a single Hilbert tensor, but a family of them. One may ask 
whether, among all these possibilities, there is a preferred choice for the Hilbert tensor. The answer is in the 
positive and it consist in determining the density weights $n_{{}_{\!A}}$ out of the physical dimensions of the 
fields. This issue is discussed in section \ref{sci}.

\subsection{Improvements}
\label{impr}
An improvement of the energy-momentum tensor has been already defined in the introduction as the addition of a
functional of the fields with identically vanishing divergence. An example of such operation is the addition of the
term $\partial_\sigma F_A^{\nu[\mu\sigma]}$ to the canonical energy-momentum tensor in (\ref{belinfdef}) to obtain the
Belinfante energy-momentum tensor. If we already have a symmetric energy-momentum tensor, improvements preserving this 
condition must be \cite{henn1,henn2} of the form $\partial_{\alpha\beta}Y^{\mu\alpha\nu\beta}$ with
$Y^{\mu\alpha\nu\beta}$ having the symmetries of the Riemman tensor, that is, antisymmetric
in $\mu\alpha$, antisymmetric in $\nu\beta$, and symmetric under the exchange $\mu\alpha\leftrightarrow \nu\beta$.

A mechanism to improve symmetric energy-momentum tensors is provided \cite{Deser:1970hs} by the covariantization 
method: one may add to the covariantized Lagrangian a new term which vanishes in the flat space limit. Supose we call
${\cal L}_{\!d}(g,\phi)$ this new addition, such that ${\cal L}_{\!d}{}_{|_{g\to\eta}} =0$, which also implies
$[{\cal L}_{\!d}]_A{}_{|_{g\to\eta}} =0\,, \ \forall \phi^A$. The Noether identity (\ref{noethidcov}), 
in the flat space limit, becomes
\beq
\partial_\mu(\frac{\delta {\cal L}_{\!d}}{\delta g_{\mu\nu}}{}_{|_{g\to\eta}})=0\,,
\label{deserimpr}
\eeq
identically. Thus the addition to an energy-momentum tensor of a term\footnote{Although ${\cal L}_{\!d}{}_{|_{g\to\eta}} =0$, the quantities $\frac{\delta {\cal 
L}_{\!d}}{\delta g_{\mu\nu}}{}_{|_{g\to\eta}}$ need not vanish.} proportional to $\frac{\delta {\cal 
L}_{\!d}}{\delta g_{\mu\nu}}{}_{|_{g\to\eta}}$ is an improvement that preserves the symmetry condition. 
An application of this procedure will be seen in subsection \ref{masslesscalar}.

\section{Scale and conformal invariance in classical field theory for $d>2$}
\label{sci}
\subsection{General conditions}
\label{ci}
In this section we consider Poincar\'e invariant Lagrangians ${\cal L}_{\!f}$ in a $d$-dimensional Minkoskwi 
spacetime, $d>2$, with $c$ and $\hbar$ as universal constants. The condition for a given spacetime vector field 
$\epsilon^\mu$ to generate a Noether symmetry in flat
spacetime can be read off from comparishon of (\ref{noethcond}) and (\ref{noethidRos2}): it requires that
$T^{\mu\nu} (\partial_\mu \epsilon_\nu+\partial_\nu \epsilon_\mu)$ be a divergence.
Now we see that Hilbert tensor is more fundamental that Belinfante's for exa\-mining the conditions
to have flat spacetime symmetries beyond Poincar\'e's, because it is with the Hilbert tensor that
we control the existence of new symmetries.
We will explore this possibility of new symmetries for the case of conformal transformations. Vector fields
$\epsilon^\mu$ generating conformal transformations\footnote{Notice that here we only analyze spacetime
conformal transformations, always within the active view. Sometimes the same language is used for Weyl
transformations, which are not spacetime transformations. Weyl symmetries, which can be rigid or gauge, are
dealt with in \cite{Forger:2003ut}.}
in flat spacetime are the solutions for $\partial_\mu \epsilon_\nu+\partial_\nu \epsilon_\mu$ to be
proportional to
$\eta _{\mu\nu}$ (conformal Killing vectors of flat spacetime). For dimensions of spacetime $d>2$ they are
\beq
\epsilon^\mu = a x^\mu +
\frac{1}{2} (x\cdot x) b^\mu - (x\cdot b) x^\mu\,,
\label{epsconf}
\eeq
($a$ is the infinitesimal parameter for scale transformations and $b^\mu$ are the infinitesimal
parameters for a special conformal trasformation).

\subsubsection{Scale invariance}
\label{Scaleinvariance}
Every Poincar\'e invariant Lagrangian ${\cal L}_{\!f}$ not containing dimensional parameters implements scale 
invariance as a classical Noether symmetry. Indeed, the scale transformation $\delta_{\!s}$ for a field --or field 
component--
$\phi^A$ is (we drop an infinitesimal constant parameter form the transformation)
\beq\delta_{\!s} \phi^A = x^\mu\partial_\mu\phi^A + d_{{}_{\!A}} \phi^A
\label{sc}
\eeq (the first term
originates from the active view of spacetime transformations. No sum for $A$ in the last term), where $d_A$ is its 
dimension in units of mass. If, and only if, there are no dimensional parameters, ${\cal L}_{\!f}$ transforms as
\beq
\delta_{\!s} {\cal L}_{\!f} = x^\mu\partial_\mu {\cal L}_{\!f} + d \,{\cal L}_{\!f} =
\partial_\mu(x^\mu{\cal L}_{\!f})
\label{scalinv}
\eeq
 and hence it is a Noether symmetry.

\vspace{4mm}

Note that the scale trasformation (\ref{sc}) can be understood as the flat spacetime limit of a
diffeomorphism transformation when $\epsilon^\mu=x^\mu$ ($x^\mu$ are Cartesian coordinates, for which the
flat metric is expressed as $\eta_{\mu\nu}$). To make this connection, one asks for
\beq\delta_{\epsilon}\phi^A = \delta_{\epsilon}^{(0)}\phi^A + n_{{}_{\!A}} \phi^A\partial_\mu\epsilon^\mu
\label{sc2}
\eeq
($\delta_{\epsilon}^{(0)}$ is the Lie derivative for a field without density behaviour) to agree with (\ref{sc}) when 
$\epsilon^\mu=x^\mu$. In fact, for such an $\epsilon^\mu$, we have, from (\ref{gaugetr}),
$\delta_{\epsilon}^{(0)}\phi^A = x^\mu\partial_\mu \phi^A + R^{(0)A\nu}_\nu$, and, using (\ref{simantisim}),
$R^{(0)A\nu}_\nu = \frac{1}{2} {\cal Q}^{(0)A \mu\nu}_B\eta_{\mu\nu}\,\phi^B$. The computation of this piece, ${\cal 
Q}^{(0)A \mu\nu}_B\eta_{\mu\nu}$, depends on the number of covariant and contravariant indices of the
field. For a field $\phi^A$ of the type
$P^{\mu_1 \mu_ 2 \ldots \mu_r}_{\nu_1 \nu_ 2 \ldots \nu_s}$, one has $\frac{1}{2} {\cal 
Q}^{(0)A\mu\nu}_{B}\eta_{\mu\nu} = (s-r) \delta_B^A$ and thus (\ref{sc}) and (\ref{sc2}) are the same
if one defines the density weight of the field $P^{\mu_1 \mu_ 2 \ldots \mu_r}_{\nu_1 \nu_ 2 \ldots \nu_s}$ as
\beq
n_{\!{}_P} = \frac{d_{\!{}_P}+r-s}{d}\,.
\label{nA}
\eeq
Note that the choice of the density weights as determined by (\ref{nA}) dictates a preferred choice for
the Hilbert tensor. We will adress this important issue in subsection \ref{pref}.

\vspace{4mm}

As examples of the application of (\ref{nA}), the scalar field, of dimension
$d_\phi = \frac{d-2}{2}$ has a density weight ${n}_\phi = \frac{d-2}{2\,d}$, and the Maxwell field, of
dimension $d_{{}_{\!A_{\!\mu}}} = \frac{d-2}{2}$ has a density weight ${n}_{{}_{\!A_{\!\mu}}} =
\frac{d-4}{2\,d}$. 

\subsubsection{Conformal invariance}
\label{scvcf}
Similarly to the scale transformations, special
conformal transformations $\delta_{\!c}$ can be read as the flat spacetime limit of diffeomorphisms with
$\epsilon^\mu =
\frac{1}{2} (x\cdot x) b^\mu - (x\cdot b) x^\mu$.
Using (\ref{simantisim}), one gets
\beq
\delta_{\!c} \phi^A = \epsilon^\mu\partial_\mu \phi^A
 -x_\mu b_\nu \, {\cal S}^{A\mu\nu}_{B} \phi^B- (x\cdot b) \,d_{\!{}_A} \phi^A\,,
\label{cf}
\eeq
where $d_{\!{}_A}$ has been obtained from (\ref{nA}). Equation (\ref{cf}) appears in \cite{Gross:1970tb} --
se also \cite{wess}-- by consistency requirements of the transformations with the algebra of the conformal group. Here 
it is a consequence of the flat spacetime limit of diffeomorphisms associated with conformal transformations, with the 
key use of (\ref{nA}).
\vspace{4mm}

For conformal Killing vectors (\ref{epsconf}), one has
$\partial_\mu \epsilon_\nu+\partial_\nu \epsilon_\mu= 2 (a - (x\cdot b))\eta _{\mu\nu}$ and thus, to generate a 
Noether symmetry, we need
\beq(a - (x\cdot b))\,T = {\rm divergence}\,,
\label{condscalconf}
\eeq
where $T$ is the trace $T^\mu_{\ \mu}$. 

Thus the condition for scale
invariance is that $T$ must be a divergence. Tis condition can be rephrased as the requirement that the e.o.m. for the 
``Lagrangian'' $T$ vanish identically, that is, $[T]=0$ for any field configuration. 

The condition for special conformal invariance is that $x^\rho\,T$ to be a divergence, that is, it requires 
$[x^\rho\,T]=0$ identically. Allowing --just for simplicity-- dependencies up to second spacetime derivatives, 
one has 
\beq
[x^\rho\,T]=x^\rho\,[T] - \frac{\partial\,T}{\partial\phi_{,\rho}}
+2\partial_\mu(\frac{\partial\,T}{\partial\phi_{,\rho\mu}})=: x^\rho\, G(\phi,\partial\phi,\partial^2\phi)+
F(\phi,\partial\phi,\partial^2\phi) =0
\label{specconf}
\eeq
for any field configuration. Considering translations $x^\rho\to x^\rho+ \lambda^\rho$ we can easily convince ourselves
that this last equation implies $G(\phi,\partial\phi,\partial^2\phi)=0$ and $F(\phi,\partial\phi,\partial^2\phi)=0$
separately. In particular this means that $[T]=0$. We conclude that the assumption of special conformal
invariance, $[x^\rho\,T]=0$ identically, already guarantees scale invariance, $[T]=0$ identically, and is
thus equivalent to full conformal invariance\footnote{This result is also obtained by examining the algebra of the 
Poincar\'e plus conformal trasformations.}.

When scale invariance is realized, $T =\partial_\mu D^\mu$ for some $D^\mu$. Plugging this relation into 
(\ref{specconf}) gives $[D^\mu]=0$ which means that $D^\mu$ must be in its turn a divergence for full conformal 
invariance to be realized. This is equivalent to the requirmeent for $T$ to be a double divergence, that is, 
there must 
exist $C^{\alpha\beta}$ such that $T= \partial_{\alpha\beta}C^{\alpha\beta}$. Notice that we can always take 
$C^{\alpha\beta}$ symmetric, which we do in the following. If these quantities
do exist, there is a standard procedure \cite{Polchinski:1987dy} to improve the energy-momentum tensor so that it will
explicitely exhibit the tracelessness condition. In $d>2$ spacetime dimensions, one defines
\bea Y^{\mu\alpha\nu\beta}&=&
\frac{1}{d-2}\Big(\eta^{\mu\nu}C^{\alpha\beta}+\eta^{\alpha\beta}C^{\mu\nu}-\eta^{\alpha\nu}C^{\beta\mu}
-\eta^{\beta\mu} C^{\alpha\nu}\Big)
\nonumber\\&-&\frac{1}{(d-1)(d-2)}\Big(\eta^{\mu\nu}\eta^{\alpha\beta}-\eta^{\alpha\nu}\eta^{\beta\mu}\Big)C^{\rho}_{\
\rho}\,.
\label{improv}\eea
$Y^{\mu\alpha\nu\beta}$ is the only structure, made with the Minkowski metric and $C^{\alpha\beta}$, with the 
symmetries of the Riemman tensor, and with the property
$\partial_{\alpha\beta}(Y^{\mu\alpha\nu\beta}\eta_{\mu\nu}) = \partial_{\alpha\beta}C^{\alpha\beta}$.
Then the improved tensor
\beq\fbox{$ \displaystyle
T_{\cal C}^{\mu\nu}:= T^{\mu\nu} - \partial_{\alpha\beta}Y^{\mu\alpha\nu\beta}
$}\eeq
is traceless. Equation (\ref{noethidRos2}) now becomes
\beq
[{\cal L}_{\!f}]_A \delta_{\!\epsilon}\phi^A-\partial_\mu\Big([{\cal L}_{\!f}]_A R^{A\mu}_\rho\epsilon^\rho - T_{\cal
C}^{\mu\sigma} \eta_{\sigma\rho}\epsilon^\rho\Big) =0\,.
\label{noethidRos5}
\eeq
with $\epsilon^\rho$ the vector field for conformal transformations
(\ref{epsconf})\,. Note that
$T_{\cal C}^{\mu\nu}-[{\cal L}_{\!f}]_A R^{A\mu}_\rho\eta^{\rho\nu}$ is an improvement of Belinfante's tensor. In fact the conserved current in (\ref{noethidRos5}) is 
$J^\mu= (T_{\!b\, \nu}^{\mu}-\partial_{\alpha\beta}Y^{\mu\alpha\sigma\beta}\eta_{\sigma\nu})\epsilon^\nu$, out of which, after moving to phase space, one can obtain the canonical generators of scale and conformal transformations.

In conclusion, scale invariance is equivalent to $T$ --the trace of the Hilbert tensor-- being a divergence, $T=\partial_\mu D^\mu$,
whereas special conformal invariance, which already implies scale invariance and thus full conformal invariance, is
equivalent to $T$ being a double divergence, $T= \partial_{\alpha\beta}C^{\alpha\beta}$, which is a much more 
restrictive condition.

\subsection{Selecting the Hilbert tensor}
\label{pref}
Consider again (\ref{noethidRos4}) and apply it to the scale invariance property of a Poincar\'e invariant Lagrangian 
without dimensional parameters. For $\delta_{\!\epsilon}\phi^A$ to become the scale transformation 
$\delta_{\!c}\phi^A$ in (\ref{sc}) we need the satisfaction of (\ref{nA}), which means that the Hilbert 
tensor present in 
(\ref{noethidRos4}) is in fact $T_{\mu\nu}^{(n)}$, associated with a specific covariantization within the family 
introduced in \ref{variety}. We infer that The Hilbert tensor becomes completely determined --up to 
improvements-- when scale invariance is implemented. Now we will compute the trace of this tensor for a scale 
invariant theory. Let us write (\ref{noethidRos4}) for $\epsilon^\mu = x^\mu$,
\beq
[{\cal L}_{\!f}]_A \delta_{\!s}\phi^A+\partial_\mu( T_{\!b}^{\mu\sigma} \eta_{\sigma\rho}\,x^\rho) -T^{(n)} =0\,.
\label{Tn}\eeq
We can isolate the trace $T^{(n)}$ from (\ref{Tn}) and obtain, using (\ref{scalinv}),
\bea
T^{(n)}&=& \partial_\mu\Big(x^\mu{\cal L}_{\!f} -  \frac{\partial
{\cal L}_{\!f}}{\partial \phi^A_{,\mu}}\delta_{\!s}\phi^A +  T_{\!b}^{\mu\sigma}
\eta_{\sigma\rho}\,x^\rho\Big)=
\partial_\mu\Big(x^\rho( T_{\!b}^{\mu\nu}-{\hat T}^{\mu\nu})\eta_{\nu\rho}  - \frac{\partial
{\cal L}_{\!f}}{\partial \phi^A_{,\mu}} d_{{}_{\!A}} \phi^A\Big)\nonumber\\
&=& \partial_\mu\Big(x^\rho(\partial_\sigma F^{\nu\mu\sigma})\eta_{\nu\rho}  - \frac{\partial
{\cal L}_{\!f}}{\partial \phi^A_{,\mu}} d_{{}_{\!A}} \phi^A\Big)= \partial_\sigma\Big(
F^{\nu\mu\sigma}\eta_{\nu\mu}- \frac{\partial
{\cal L}_{\!f}}{\partial \phi^A_{,\sigma}} d_{{}_{\!A}} \phi^A\Big)\nonumber\\
&=&\partial_\sigma\Big( \frac{\partial
{\cal L}_{\!f}}{\partial \phi^A_{,\mu}}( {\cal S}^{A \sigma\nu}_B\eta_{\mu\nu}\phi^B - \delta^\sigma_\mu
d_{{}_{\!A}} \phi^A  ) \Big) = \partial_\sigma\Big( \frac{\partial
{\cal L}_{\!f}}{\partial \phi^A_{,\mu}}\eta_{\mu\nu}( {\cal S}^{A \sigma\nu}_B -
d_{{}_{\!A}}\delta^A_B\,\eta^{\sigma\nu}     )\phi^B \Big)\,,
\label{prefhr}\eea
where we have used (\ref{scalinv}) and the definitions (\ref{naive}),\, (\ref{deff}) and (\ref{belinfdef}), and the antisymmetry
properties of $F^{\nu\mu\sigma}$. From the first equality in (\ref{prefhr}) we see that as expected
$T^{(n)}$ is
a divergence, which is the condition for scale invariance. 

\vspace{4mm}

Once this tensor is determined, the fate of conformal invariance is sealed and its  
possible additional realization can be further investigated. In fact, (\ref{prefhr}) shows explicitely the new 
condition for
implementing full conformal invariance in the presence of scale invariance: for $T^{(n)}$ to be a double divergence 
(which is the requirement found in subsection \ref{scvcf}) we need
\beq\fbox{$ \displaystyle\frac{\partial
{\cal L}_{\!f}}{\partial \phi^A_{,\mu}}\eta_{\mu\nu}\Big( {\cal S}^{A \sigma\nu}_B -
d_{{}_{\!A}}\delta^A_B\,\eta^{\sigma\nu}     \Big)\phi^B = {\rm divergence}\,.
\label{confcond}$}\eeq
This result can be stated as the following

\underline {\bf Proposition}.\quad{\it A Poincar\'e invariant Lagrangian ${\cal L}_{\!f}$ in
$d$-dimensional Minkowski spacetime, with $d>2$, with no
dimensional parameters --and thus implementing scale inva\-riance--, is conformal invariant if and only if
expression (\ref{confcond}) is fulfilled.}

\vspace{4mm}

This equation (\ref{confcond}) was obtained in \cite{Callan:1970ze} (see section 5.2 in that paper) on the basis of an 
analysis of the conditions for $\delta_c$ in (\ref{cf}) to be a Noether transformation. The difference with our 
approach is that we can specify the Hilbert tensor which contains in its 
trace the information as to whether a scale invariant theory realizes conformal invariance. It is worth noticing that 
this Hilbert tensor is not the Belinfante tensor --though they differ in terms vanishing on shell-- and it is only 
this Hilbert tensor that which becomes traceless --if conformal invariance holds-- after the improvement devised at 
the end of subsection \ref{scvcf}.

\subsection{Examples}
\subsubsection{Massless real scalar field with $\phi^m $ interaction in $d>2$ dimensions}
\label{masslesscalar}
The Poincar\'e invariant Lagrangian is
${\cal L}_{\!f}= \frac{1}{2} \partial_\mu \phi \partial_\nu \phi\,  \eta^{\mu\nu} + \Lambda \phi^m\,.
$ The dimension of the field is $d_\phi =   \frac{d-2}{2}$. According to subsection \ref{Scaleinvariance}, scale 
invariance is guaranteed as long as $\Lambda$ is dimensionless, which is equivalent to  $m= \frac{2\,d}{d-2}$. Its 
integer solutions are $(d,m) = (6,3),(4,4),(3,6)$. It is now easy to check that  conformal invariance holds at any 
dimension. Indeed, the kinetic term satisfies (\ref{confcond}) --because ${\cal S}^{A \sigma\nu}_B$ vanishes whereas 
$\phi\, \partial_\mu \phi$ is a divergence
 -- and so does trivially the selfinteraction term --because since it has no
derivatives, it does not contribute to (\ref{confcond}). The rhs of (\ref{confcond}) is actually 
$\partial_\mu(-\frac{d-2}{4} \eta^{\mu\sigma}\phi^2)$. Constructing the
improved, traceless, energy-momentum tensor, is, after (\ref{improv}), straightforward. As explained in subsection 
\ref{impr}, an alternative procedure to introduce the improvement is the addition to the covariantized Lagrangian of a 
term proportional to $\sqrt{| g |} R \phi^2$, see \cite{Deser:1970hs}, taking into account that $\frac{\delta (\sqrt{| 
g |} R \phi^2)}{\delta g^{\mu\nu}}|_{g\to\eta} = 
\partial_{\alpha\beta}\Big((\eta_{\mu\nu}\eta^{\alpha\beta}-\delta_\mu^\alpha \delta_\nu^\beta)\phi^2\Big)$.

\vspace{4mm}

The extension to $O(N)$ symmetry, with interaction term $\Lambda ({\vec\phi}^2)^\frac{m}{2}$ is immediate and one 
can check that for dimensionless $\Lambda$, both scale and conformal invariance are realized at any spacetime 
dimension $d>2$.

\subsubsection{Pure Maxwell or YM theory}

Let us apply (\ref{confcond}) to the Maxwell Lagrangian, ${\cal L}_{{}_{\!m\!a\!x}}=-\frac{1}{4}F_{\mu\nu}F^{\mu\nu}$, 
with $F_{\mu\nu} 
:=\partial_{\mu} A_{\nu}- \partial_{\nu} A_{\mu}$, for which $d_{{}_{\!A_{\!\mu}}} = \frac{d-2}{2}$. Expression 
(\ref{confcond}) becomes
\bea &&\frac{\partial
{\cal L}_{{}_{\!m\!a\!x}}}{\partial A_{\rho,\mu}}\eta_{\mu\nu}
\Big( {\cal S}^{(\lambda) \sigma\nu}_{(\rho)} -
\frac{d-2}{2}\delta^\lambda_\rho\,\eta^{\sigma\nu}     \Big)A_\lambda =
-F^{\mu\rho}\eta_{\mu\nu}
\Big(\delta^\nu_\rho \eta^{\lambda\sigma}- \delta^\sigma_\rho \eta^{\lambda\nu} -
\frac{d-2}{2}\delta^\lambda_\rho\,\eta^{\sigma\nu}\Big)A_\lambda\nonumber\\
&&=-F^{\mu\rho}\Big(\eta_{\mu\rho} A^\sigma - \delta^\sigma_\rho
A_\mu - (1+ \frac{d-4}{2}) \delta^\sigma_\mu A_\rho\Big) =
\frac{d-4}{2} F^{\sigma\rho}\, A_\rho\,,
\label{maxconf}\eea
(where we have used (\ref{vecrep})) which is only a divergence
--actually it vanishes-- for $d=4$. Thus pure Maxwell, which is
scale invariant at any spacetime dimension, only becomes conformal
invariant at four dimensions. 

One can easily check that scalar
electrodynamics for $d=4$ also satisfies (\ref{confcond}). On the
other hand, For pure YM, due to the presence of the
selfinteraction terms, scale and/or conformal invariance are only
achieved at four dimensions (it is the spacetime dimension for which the coupling constant is dimensionless). The 
proof repeates the calculations done for the Maxwell case but with a new summation index over the dimension of the 
gauge group.

\subsubsection{$p$-forms}
\label{pfroms}
Let us first consider the free theory of $2$ forms, with gauge field ${\bf B} = \frac{1}{2} B_{\mu\nu}dx^\mu\wedge 
dx^\nu$, with $B_{\mu\nu}$ antisymmetric. The field strenght is 
$H_{\mu\nu\sigma}= \partial_\mu B_{\nu\sigma}+ \partial_\nu B_{\sigma\mu}+\partial_\sigma B_{\mu\nu}$ and the 
Lagrangian is the kinetic term ${\cal L}_{{}_{\!B}}= \frac{1}{6}H_{\mu\nu\sigma}H^{\mu\nu\sigma}$. To compute the lhs 
of (\ref{confcond}) we note that the representation of the Lorentz algebra defined in 
(\ref{rversuss}) is now\footnote{We apply the practical convention that any object with indices that saturate those of 
$B_{\mu\nu}$ must be taken antisymmetric in such indices. This is the reason of the factor $\frac{1}{2}$.} ($A= 
[\mu\nu],\ B= [\alpha\beta]$)
$$
{\cal S}^{[\alpha\beta]\sigma\rho}_{[\mu\nu]} = \frac{1}{2}\Big(\eta^{\alpha\sigma}\rho^\beta_\nu\delta^\rho_\mu
- \eta^{\alpha\rho}\delta^\beta_\nu\delta^\sigma_\mu+ \eta^{\beta\sigma}\delta^\alpha_\mu\delta^\rho_\nu- 
\eta^{\beta\rho}\delta^\alpha_\mu\delta^\sigma_\nu-(\mu\leftrightarrow\nu)\Big)\,,$$
and (\ref{confcond}) becomes
\bea &&
\frac{\partial {\cal L}_{{}_{\!B}}}{\partial B_{\mu\nu,\lambda}}\eta_{\lambda\rho}\Big( {\cal 
S}^{[\alpha\beta]\sigma\rho}_{[\mu\nu]}- \frac{d-2}{2} \delta^{[\alpha\beta]}_{[\mu\nu]} \eta^{\sigma\rho}    \Big) 
B_{\alpha\beta}\nonumber\\
&=& H^{\lambda\mu\nu}\eta_{\lambda\rho}\Big( - \delta_\mu^\sigma B_{\lambda\nu} - \delta_\nu^\sigma B_{\mu\lambda}  
-(2+ \frac{d-6}{2} )  \delta_\lambda^\sigma B_{\mu\nu}          \Big)\nonumber\\
&=& -\frac{d-6}{2} H^{\sigma\mu\nu}B_{\mu\nu}\,,
\label{bconf}\eea
(where we have used in the second and third equalities the full antisymmetry of $H^{\lambda\mu\nu}$) which is only a 
divergence --in fact it vanishes-- for $d=6$. It is not difficult to identify a pattern in 
the expressions (\ref{maxconf}) and (\ref{bconf}) which allows for a generalization to $p$-forms. The result for the 
$p$-form free theory is that conformal invariance is only achieved for $d=2(p+1)$ dimensions. Since, according to 
(\ref{nA}), the density weight of a $p$-form field is $n_p= \frac{d-2(p+1)}{2\,d}$, one can deduce from the generalization of (\ref{bconf}) that the Hilbert 
tensor $T_{\mu\nu}^{(0)}$ obtained from the minimal covariantization will become tracelessness for $d=2(p+1)$. 

\subsubsection{Interaction term with scale invariance but not conformal invariance}
\label{counter}
Consider the interaction term of a scalar field with a vector field  ${\cal L}_{int} = \Lambda \eta^{\mu\nu} 
(\partial_\mu A_\nu) \phi^m$. It breaks $U(1)$ gauge as well as rigid invariance\footnote{In \cite{Callan:1970ze} the 
case with an interaction term of the type $(\partial_\mu A^\mu)^2$ is considered.} and we can even take the scalar 
field real. We assume the standard dimensions for the fields $A_\mu$ and $\phi$. Scale invariance is guaranteed with 
dimensionless $\Lambda$, which means $m= \frac{d}{d-2}$, whose 
integer solutions are $(d,m)= (3,3),(4,2)$. As regards conformal invariance, the lhs of (\ref{confcond}) becomes in 
this case $\Lambda \frac{d}{2} A^\sigma \phi^m$,  which is not a divergence. Thus the model admits scale invariance 
but not conformal invariance. This model though does not lead to a unitary quantum field theory. In this sense, the 
issue \cite{Polchinski:1987dy} as to whether there exists a $d=4$ unitary quantum field theory realizing scale 
invariance but not conformal invariance remains unsettled\footnote{The assumptions made in \cite{Gross:1970tb} exclude 
our example.}.

\subsubsection{Abelian Chern-Simons theory in $d= 2n+1$ dimensions}
\label{ch-s}

The Lagrangian density is
\beq
{\cal L}_{{}_{\!C\!S}}^{2n+1}= \epsilon^{\mu_{0}
\mu_1...\mu_{2n}}A_{\mu_{0}}F_{\mu_{1}\mu_{2}} \ldots 
F_{\mu_{2n-1}\mu_{2n}}\,.
\eeq
This Lagrangian, which is metric independent, is already diffeomorphism invariant and therefore scale and conformal 
invariance are trivially realized as part of it. The dimension of the field is 
$d_{{}_{\!A_{\!\mu}}} = 1$ and the corresponding density weight, from (\ref{nA}), is $n_{\!A_{\!\mu}} = 0$, which is 
consistent with the fact that the Hilbert tensor vanishes identically ($n_{\!A_{\!\mu}}\neq 0$ would have 
suggested a rewriting of the "covariantized" Lagrangian which would have allowed to define along the lines of 
subsection \ref{variety}, see (\ref{newros}), a nonvanishing 
Hilbert tensor $T_{\mu\nu}^{(n)}$ even when $T_{\mu\nu}^{(0)}$ was vanishing). One can also check that the 
lhs of (\ref{confcond}) vanishes identically. Equation (\ref{noethidRos4}) remains valid but without the last term of 
the lhs. One directly identifies from this expression the Noether current for diffeomorphism invariance as 
$J^\mu= T^{\mu}_{\!b\, \nu}\,\epsilon^\nu$, 
where, being zero the Hilbert tensor, the Belinfante tensor can be extracted from 
(\ref{tbr}) as $T^{\mu}_{\!b\, \rho}=-[{\cal L}_{\!f}]_A R^{A\mu}_\rho$, which makes equation (\ref{noethidRos4}) to 
become the Noether identity (\ref{noethid1}). The Belinfante tensor vanishes on shell but it is 
capable of providing in the canonical picture for the generators of the gauge symmetry of diffeomorphism invariance.

\section{Classical field theory with gravity}
Consider a matter covariant Lagrangian ${\cal L}_m(g, \partial g, \phi,\partial\phi)$ (where $g$ is the
metric field and $\phi$ the matter fields). Following Hilbert's prescription, we define the matter
energy-momentum tensor (Hilbert) density\footnote{it is a tensor density of weight $1$ because the
Lagrangian is a scalar density.} $T^{\mu\nu} = - 2 \frac{\partial {\cal L}_m}{\partial
g_{\mu\nu}}$. Since ${\cal L}_m$ is a scalar density, we can write the Noether identities with respect to
diffeomorphism invariance for ${\cal L}_m$. We will use the manifestly covariant form
of the Noether identities discussed on subsection \ref{covnoeth}. We have, for the variations (see
(\ref{gaugetrcov})),
$$\delta_{\!\epsilon} g_{\mu\nu}=\nabla_\mu\epsilon_\nu+ \nabla_\nu\epsilon_\mu\,,\ \ \delta_{\!\epsilon}
\phi^A=U^A_\rho\epsilon^\rho + R^{A\mu}_\rho \nabla_\mu\epsilon^\rho\,,
$$
and the the Noether identities (\ref{noethidcov}), now for fields $g_{\mu\nu}$ and $\phi^A$, become
\beq \nabla_\mu T^\mu_{\ \nu} = -[{\cal L}_m]_A U^A_\nu + \nabla_\mu\Big( [{\cal L}_m]_A R^{A\mu}_\nu
\Big)\,.
\label{roscurv}
\eeq
Saturating with $\epsilon^\nu$ we obtain an expression formally identical to (\ref{noethidRos2}),
\beq
[{\cal L}_m]\delta_{\!\epsilon} \phi^A + \partial_\mu\Big( (T^\mu_{\ \nu} - [{\cal L}_m]_A R^{A\mu}_\nu
)\epsilon^\nu \Big) - T^\mu_{\ \nu}\nabla_\mu\epsilon^\nu=0\,,
\label{roscurv2}
\eeq
but now in curved, dynamical spacetime. More important is to notice that (\ref{roscurv}) tells that
$T^\mu_{\ \nu}$
is a covariantly conserved tensor density as long as the matter fields satify the e.o.m., and this result
has no relation whatsoever with the Einstein equations --we are only considering the matter Lagrangian 
(had we considered the addition of the purely gravitational Einstein-Hilbert Lagrangian, the
Einstein e.o.m.$\!$ will also imply independently that $T^\mu_{\ \nu}$ is covariantly conserved). This point, also 
made in \cite{Gotay}, is basically known since the early years of GR \cite{Hilbert} but sometimes gets blurred. 
Contrary to some popularizations, it is not that
the choice for the Einstein tensor density --the gravity side of the Einstein equations-- is restricted
because one
{\it wants to have} a covariantly conserved matter energy-momentum tensor, but because one already {\it has}
it. Of
course, when the e.o.m.$\!$ are derived from a variational principle there is no restriction any more because 
any purely gravitational metric Lagrangian ${\cal L}_{grav}$ will yield, as Noether identities associated with
diffeomorphism invariance, the covariant conservation of $\frac{\delta {\cal L}_{grav}}{\delta g_{\mu\nu}}$.

\subsection{Field theory in nondynamical curved spacetime}
One could decide to freeze the background, let's name it $\bar g$, and to make it non-dynamical. This
describes matter in a fixed background without taking into account the backreaction induced by the matter on
it. There is a case, though, where the backreaction is indeed included, which is when, in addition to the
satisfaction of the matter e.o.m., $\bar g$ is solution of the e.o.m. for ${\cal L}_{grav} + {\cal L}_m$.

Now the Lagrangian ${\cal L}_m$ becomes a functional of the matter fields only,
${\cal L}_{\bar m} = {\cal L}_m{}_{|_{g\to\bar g}}$. Equations (\ref{roscurv}) and (\ref{roscurv2}) are still
applicable with the only circumstance of sending $g\to\bar g$. Thus for any matter Lagrangian in any fixed background,
there is covariant conservation of the Hilbert tensor. Is this conservation related to any Noether symmetry? The
answer in general is no\footnote{Since the background is fixed, there is no longer the gauge symmetry of 
diffeomorphism invariance. A residual, non-gauge, subroup of diffeomorphisms can still realize Noether symmetries, 
as we shall see.}, because, except for some special cases, there is no way to extract from this tensor density a 
conserved\footnote{Conserved or covariantly conserved is
the same for a vector density, with the standard Riemmanian connection.} vector density --to play the role of the
Noether current. The exceptions: first, when the background has Killing vectors. If $\xi^\mu$ is a Killing vector for 
the background  $\bar g_{\mu\nu}$ then the last term in (\ref{roscurv2}) vanishes for $\epsilon^\mu\to\xi^\mu$ and we 
end up with the Noether conserved current
$J^\mu = (T^\mu_{\ \nu} - [{\cal L}_{\bar m}]_A R^{A\mu}_\nu) \xi^\nu$\,. Another exception is the case when 
$\epsilon^\mu$ is such that $T^\mu_{\ \nu}\nabla_\mu\epsilon^\nu $ happens to be a divergence (see (\ref{roscurv2})). 
This is reminiscent of our explorations in subsection \ref{ci} concerning conformal symmetry in flat spacetime. In 
fact, for a conformal Killing vector the condition is that $(\nabla_\mu\epsilon^\mu) tr(T)$ must be a divergence, 
which includes the tracelessness case $tr(T)=0$, now in curved space.

The tensor $T^\mu_{\ \nu} - [{\cal L}_{\bar m}]_A R^{A\mu}_\nu$ could well receive the name of Belinfante --or
Belinfante-Rosenfeld-- tensor in nondynamical curved space, because it is the tensor that parti\-cipates in defining 
the Noether conserved current and the canonical Noether symmetry generator, if such a symmetry is realized.

\section{Conclusions}

We have presented a basic review of classical aspects of Noether symmetries, with special emphasis in the use of 
Noether identities for gauge theories. Next we have discussed in detail the role of diffferent energy-momentum tensors 
in Poincar\'e invariant field theories and explored the eventual realization of scale and conformal invariance. The 
core of this part of paper is devoted to flat spacetime.

The main points worth remarking are
\begin{itemize}
\item The introduction and use of some identities of the variational calculus, (\ref{kiyoshi}), in order
to obtain the conditions for a continuous symmetry.
\item The use of the Noether identities to find the first class structure --except for the primary
constraints in phase space-- of constraints of a gauge theory, subsections \ref{constr-noethidsect} and \ref{agen}.
\item The observation that whereas the Belinfante energy-momentum tensor is unique and it is
directly linked with the Poincar\'e generators, there is a family of Hilbert tensors, subsection 
\ref{variety}, which are the means to test whether further spacetime symmetries can be realized. All these tensors 
only differ in terms vanishing on shell.
\item The selection of a unique Hilbert tensor for Poincar\'e invariant Lagrangians realizing scale 
invariance, subsection \ref{pref}.
\item An example, subsection \ref{counter}, of a toy model that exhibits scale invariance but not conformal invariance 
at the classical level.
\item A simple expression, (\ref{confcond}), derived form the computation of the trace of the Hilbert tensor, as a 
means to check for a specific scale invariant theory whether conformal invariance is also realized.
\end{itemize}
Let us elaborate on this last point, summarizing our results. We start with Poincar\'e invariant theories in 
Minkowski spacetime. In the classical setting, scale invariance is identified by the absence of dimensional parameters 
in the Lagrangian. The correct Hilbert tensor for such a scale invariant theory is determined through 
(\ref{newros}) by using the relation (\ref{nA}) between the dimensions of the fields and their density weights upon 
covariantization. The trace of this tensor is a divergence --thus describing scale invariance--, and it is computed in 
(\ref{prefhr}). Then the additional requirement of conformal invariance is (\ref{confcond}).

\section*{Acknowledgements}
The author thanks Kiyoshi Kamimura and Joseph Polchinski for useful exchanges and to Donald Salisbury and
Kurt Sundermeyer for their comments and suggestions. Partial support form  MCYT FPA 2007-66665,
CIRIT GC 2005SGR-00564, Spanish Consolider-Ingenio 2010 Programme
CPAN (CSD2007-00042) is acknowledged. The author also thanks the hospitality of the Theoretical Physics
Group at the Imperial College London, where part of the present work was carried out, with particular
thanks to Jonathan Halliwell and Hugh Jones.

\section*{Appendix. Belinfante - Hilbert}
Here we prove the result advanced in subsection \ref{rosenfeld} concerning the relation between Belinfante
tensor and the Hilbert tensors.
Let us examine again the consequences of the Noether condition (\ref{diverg}) for the minimally
covariantized
Lagrangian ${\cal L}_{\!g} := {\cal L}_{\!g}^{(0)}$ (we eliminate superscripts for simplicity) of the flat
spacetime Lagrangian ${\cal L}_{\!f}$,
$$\delta {\cal L}_{\!g} = -\frac{1}{2} T_{\!g}^{\mu\nu} \delta g_{\mu\nu}
+ \partial_\sigma(\frac{\partial {\cal L}_{\!g}}{\partial g_{\mu\nu,\sigma}}\delta g_{\mu\nu}) + [{\cal
L}_{\!g}]_A
\delta\phi^A + \partial_\sigma(\frac{\partial {\cal L}_{\!g}}{\partial \phi^A_{,\sigma}}\delta\phi^A) =
\partial_\sigma(\epsilon^\sigma {\cal L}_{\!g})
$$
where we have used the definition (\ref{deftr}).

We can make the limit $g\to\eta$ in this expression, still keeping $\epsilon^\sigma$ arbitrary.
$T_{\!g}^{\mu\nu}$ becomes
the Hilbert tensor $T^{\mu\nu}$ associated with the covariantization. We get,
\beq -\frac{1}{2} T^{\mu\nu} \delta \eta_{\mu\nu}  + \partial_\sigma(\frac{\partial {\cal L}_{\!g}}{\partial
g_{\mu\nu,\sigma}}{\!\Big|_\eta}\delta \eta_{\mu\nu} ) + [{\cal L}_{\!f}]_A \delta\phi^A +
\partial_\sigma(\frac{\partial
{\cal L}_{\!f}}{\partial \phi^A_{,\sigma}}\delta\phi^A) = \partial_\sigma(\epsilon^\sigma {\cal L}_{\!f})
\label{twosides}\eeq
with $\delta \eta_{\mu\nu} = \eta_{\mu\rho} \epsilon^\rho_{,\nu}+ \eta_{\nu\rho} \epsilon^\rho_{,\mu}\ $
and $ \delta\phi^A = R^A_\rho \epsilon^\rho + R^{A\mu}_\rho \partial_\mu\epsilon^\rho$. Since $\epsilon^\mu$
is arbitrary, we can get different identities by equalizing the coefficients of $\epsilon^\mu,\
\partial_\sigma\epsilon^\rho$ and $\partial_{\sigma\lambda}\epsilon^\rho$ on both sides of (\ref{twosides}).
For $\epsilon^\mu$ we just get the conservation on shell of the canonical energy-momentum tensor ${\hat
T}^{\mu\nu}$, (\ref{naive}). For $\partial_\sigma\epsilon^\rho$ we obtain
\beq
- T^\sigma_{\ \rho} + {\hat T}^\sigma_{\ \rho} + [{\cal L}_{\!f}]_A R^{A\sigma}_\rho
+ \partial_\lambda\Big(2 \frac{\partial {\cal L}_{\!g}}{\partial
g_{\mu\sigma,\lambda}}{\!\Big|_\eta}\eta_{\mu\rho}+
\frac{\partial {\cal L}_{\!f}}{\partial \phi^A_{,\lambda}} R^{A\sigma}_\rho\Big) = 0 \,,
\label{parteps}
\eeq
and finally, for $\partial_{\sigma\lambda}\epsilon^\rho$ we obtain that the quantity inside the $\lambda$-derivative
in (\ref{parteps}) must be antisymmetric in the $\sigma\lambda$ indices,
that is, the quantity
\beq G^{\rho\sigma\lambda}:= 2 \frac{\partial {\cal L}_{\!g}}{\partial
g_{\rho\sigma,\lambda}}{\!\Big|_\eta}
+ \frac{\partial {\cal L}_{\!f}}{\partial \phi^A_{,\lambda}} R_{\mu}^{A\sigma} \eta^{\mu\rho}\,.
\label{interesting}
\eeq
satisfies $G^{\rho\sigma\lambda} = G^{\rho[\sigma\lambda]}$.
Substituting (\ref{simantisim}) in the last term of (\ref{interesting}), one sees that the structure of
(\ref{interesting}) is of the type
\beq
A^{(\rho\sigma)\lambda} + B^{[\rho\sigma]\lambda} = G^{\rho[\sigma\lambda]}\,,
\label{interesting2}
\eeq
and it is well known that knowledge of the coefficients $B$ completely determines the coefficients $A$ and $G$. Indeed one gets
$$A^{(\rho\sigma)\lambda} = B^{[\lambda\sigma]\rho}+B^{[\lambda\rho]\sigma}\,.$$
In our case, we obtain\footnote{This equation could in principle have been alternatively obtained by noticing that the dependences of ${\cal L}_{\!g}$ on derivatives of the metric must come form terms in ${\cal L}_{\!f}$ depending on the derivatives of the fields, since these derivatives become covariant derivatives upon covariantization.}
\beq
2 \frac{\partial {\cal L}_{\!g}}{\partial g_{\rho\sigma,\lambda}}{\!\Big|_\eta} = \frac{1}{2}\Big(\frac{\partial {\cal L}_{\!f}}{\partial \phi^A_{,\rho}} {\cal S}^{A\lambda\sigma}_B \phi^B+\frac{\partial {\cal L}_{\!f}}{\partial \phi^A_{,\sigma}} {\cal S}^{A\lambda\rho}_B \phi^B - \frac{\partial {\cal L}_{\!f}}{\partial \phi^A_{,\lambda}} {\cal Q}^{A\rho\sigma}_B \phi^B\Big)\,,
\eeq
and equation (\ref{parteps}) becomes
\bea T^{\sigma\rho} &=& {\hat T}^{\sigma\rho} + [{\cal L}_{\!f}]_A R^{A\sigma}_\mu \eta^{\mu\rho}
+ \frac{1}{2}\partial_\lambda\Big(\frac{\partial {\cal L}_{\!f}}{\partial \phi^A_{,\rho}} {\cal S}^{A\lambda\sigma}_B \phi^B+ \frac{\partial {\cal L}_{\!f}}{\partial \phi^A_{,\sigma}} {\cal S}^{A\lambda\rho}_B \phi^B+\frac{\partial {\cal L}_{\!f}}{\partial \phi^A_{,\lambda}} {\cal S}^{A\rho\sigma}_B \phi^B\Big)\nonumber\\ &=&
{\hat T}^{\sigma\rho} + [{\cal L}_{\!f}]_A R^{A\sigma}_\mu \eta^{\mu\rho}
+ \partial_\lambda(F^{\rho\sigma\lambda})
\label{bhr2}\eea
where $F^{\rho\sigma\lambda} = F^{\rho[\sigma\lambda]} = G^{\rho[\sigma\lambda]}$ are the quantities defined in (\ref{deff}). Using the definition of Belinfante's tensor, (\ref{belinfdef}), we arrive at
\beq T^{\sigma\rho}= T_{\!b}^{\sigma\rho} + [{\cal L}_{\!f}]_A R^{A\sigma}_\mu \eta^{\mu\rho}\,,
\label{bhr}
\eeq
which shows the connection between Belinfante's and the minimal --that is, obtained by the minimal covariantization-- Hilbert's tensor, anticipated in section \ref{rosenfeld}. If, however,
instead of the minimal covariantization, ${\cal L}_{\!g}^{(0)}$, we take any covariantization,
${\cal L}_{\!g}^{(0)}\to{\cal L}_{\!g}^{(n)}$, and the respective change for the variations, $\delta^{(0)}\phi^A\to
\delta^{(n)}\phi^A = \delta^{(0)}\phi^A + n_{\!A}\phi^A\partial_\mu\epsilon^\mu$, note that ${\cal S}^{A \mu\nu}_B\phi^B $ in (\ref{simantisim}) remains unchanged and, as a consequence, the Belinfante tensor in
(\ref{bhr2}),(\ref{bhr}), remains unchanged as well.

\end{document}